\documentclass[11pt,british]{article}
\usepackage[utf8]{inputenc}
\usepackage{geometry}
\usepackage[dvipsnames]{xcolor}
\usepackage{tabularx} 
\usepackage{amsmath}  
\usepackage{amssymb}
\usepackage{braket}
\usepackage{physics}
\usepackage{cite}
\usepackage{dsfont}
\usepackage{listings}
\usepackage{graphicx, wrapfig} 
\usepackage{multirow}
\usepackage[final]{hyperref} 
\hypersetup{
	colorlinks=true,       
	linkcolor=red,        
	citecolor=red,        
	filecolor=magenta,     
	urlcolor=blue         
}
\usepackage{subcaption}
\usepackage{ragged2e}
\DeclareCaptionJustification{justified}{\justifying}
\numberwithin{equation}{section}
\numberwithin{figure}{section}
\numberwithin{table}{section}
\usepackage{authblk}

\begin{document}

\title{Chirally Factorised Truncated Conformal Space Approach}
\author[1]{D.~X.~Horv{\'a}th\thanks{The authors contributed equally to this paper.}}
\author[2,3]{K.~H{\'o}ds{\'a}gi$^*$}
\author[2,4]{G.~Tak{\'a}cs}
\affil[1]{SISSA and INFN Sezione di Trieste}
\affil[2]{Department of Theoretical Physics,
Budapest University of Technology and Economics\\
H-1111 Budapest M{\H uegyetem} rkp. 3}
\affil[3]{BME-MTA Statistical Field Theory ‘Lend{\"u}let’ Research Group, 
Budapest University of Technology and Economics}
\affil[4]{MTA-BME Quantum Correlations Group (ELKH), 
Budapest University of Technology and Economics}

\date{10th April 2022}

\maketitle

\begin{abstract}
    Truncated Conformal Space Approach (TCSA) is a highly efficient method to compute spectra, operator matrix elements and time evolution in quantum field theories defined as relevant perturbations of $1+1$-dimensional conformal field theories. However, similarly to other exact diagonalisation methods, TCSA is ridden with the ``curse of dimensionality'': the dimension of the Hilbert space increases exponentially with the (square root of the) truncation level, limiting its precision by the available memory resources. Here we describe an algorithm which exploits the chiral factorisation property of conformal field theory with periodic boundary conditions to achieve a substantial improvement in the truncation level. The Chirally Factorised TCSA (CFTCSA) algorithm presented here works with inputs describing the necessary CFT data in a specified format. It makes possible much more precise calculations with given computing resources and extends the reach of the method to problems requiring large Hilbert space dimensions. In fact, it has already been used in a number of recent works ranging from determination of form factors, through studying confinement of topological excitations to non-equilibrium dynamics. Besides the description of the algorithm, a MATLAB implementation of the algorithm is also provided as an ancillary file package, supplemented with example codes computing spectra, matrix elements and time evolution, and with CFT data for three different quantum field theories. We also give a detailed how-to guide for  constructing the required CFT data for Virasoro minimal models with central charge $c<1$, and for the massless free boson with $c=1$. 
\end{abstract}

\tableofcontents

\section{Introduction}

Quantum field theories (QFTs) are fundamental not only to our understanding of the elementary particles and their interactions, but also in statistical physics and the description of condensed matter systems. In particular, $1+1$-dimensional quantum field theories play an important role in our understanding of strongly correlated many-body systems. A powerful approach is to classify first the scale-invariant QFTs, so-called conformal field theories (CFTs) \cite{1984NuPhB.241..333B} which describe fixed points of the renormalisation group flow and classify universality classes of critical behaviour in statistical systems. More general theories can then be considered as perturbations of CFTs by relevant scaling operators, many of which result in integrable models \cite{Zamolodchikov:1989hfa}, which allows exact determination of many physical properties. For a comprehensive review on this subject the reader is referred to \cite{Mussardo:2010mgq}. 

Integrable field theories, however, are only a small subclass of perturbed conformal field theories. Truncated Hamiltonian approaches are independent of integrability and therefore are applicable to general perturbed conformal field theories. The first, and still widely used application of Hamiltonian truncation to quantum field theories was the truncated conformal space approach (TCSA) to numerically study the finite volume spectra of relevant perturbations of conformal field theories, introduced by Yurov and Zamolodchikov \cite{1990IJMPA...5.3221Y}. Several variants were subsequently developed to treat a larger class of models \cite{1991IJMPA...6.4557Y,1998PhLB..430..264F,Fonseca_2001,2014JSMTE..12..010C,2015PhRvD..91b5005H,2016JHEP...10..050B,2019JHEP...07..173B}; for a recent review c.f. the paper by James et al. \cite{2018RPPh...81d6002J}. It can also be applied to boundary \cite{1998NuPhB.525..641D,2001NuPhB.614..405B} and defect problems \cite{2009JHEP...11..057K,2014NuPhB.886...93B,2014NuPhB.882..501B}. Although applications of Hamiltonian truncation are mostly dominated by 1+1 dimensional field theories, it is possible to extend the approach to higher dimensions as well \cite{2015PhRvD..91h5011R,2016PhRvD..93f5014R}. Besides the spectral problem, it enables the study of operator matrix elements in finite volume \cite{2008NuPhB.788..167P,2008NuPhB.788..209P}, and recently was also applied to investigate non-equilibrium dynamics in quantum field theories, taking into account the full quantum dynamics \cite{2016NuPhB.911..805R,2017PhLB..771..539H,2018ScPP....5...27H,2018PhRvL.121k0402K,2019JHEP...08..047H,2019PhRvA.100a3613H,2021PhRvD.104b1702K}. A recent alternative approach to Hamiltonian truncation is Lightcone Conformal Truncation \cite{2016JHEP...07..140K} which, in contrast to the TCSA, works in infinite volume; a pedagogical introduction can be found in \cite{2020arXiv200513544A}.

The above range of applications explains the interest in developing efficient algorithms for TCSA which make possible more precise computations and extend its range of applicability. Starting with the simple implementations \cite{1991CoPhC..66...71L}, several people and groups developed algorithms for TCSA in the course of their work; however, so far very few information has been published on these developments, which presents a hard barrier for anyone who wishes to apply these methods. 

The TCSA works in finite volume, and uses an energy cut-off to make the Hilbert space finite dimensional. The presence of the cut-off introduces an approximation, which needs to be controlled carefully to understand the precision of the results. In addition, it is very important to go beyond naive implementations of the original idea to address models where the Hilbert space grows rapidly with the cut-off, which is the case e.g. in models of great physical relevance with $SU(2)$ current algebra \cite{2013NuPhB.877..457B,2015NuPhB.899..547K,2016PhRvD..94d5003A}. 

The cut-off dependence can be theoretically computed, and thereby suppressed, by renormalisation group (RG) methods \cite{2006hep.th...12203F,2008JSMTE..03..011F,2011arXiv1106.2448G,2015PhRvD..91h5011R}, which rely on the validity of perturbation theory for high energy modes. However, in order to use the RG improvements it is necessary to reach sufficiently high values of the cut-off, which can be achieved by the application of numerical renormalisation group (NRG) methods \cite{2007PhRvL..98n7205K,2009PhRvL.102i7203K}. Although the NRG approach is efficient, it comes with the price of introducing further approximations which are not under theoretical control.

Here we report an algorithmic development with efficiency similar to the NRG method, which makes it possible to reach high cut-off values without introducing further approximations to the TCSA. Our Chirally Factorised Truncated Conformal Space Approach (CFTCSA) works for periodic boundary conditions and relies on the chiral factorisation of conformal field theories, which is the decoupling of left and right moving degrees of freedom. The only input the algorithm needs is a description of the Hilbert space in terms of the chiral degrees of freedom, the structure constants of the operator algebra, and so-called chiral data which are the building blocks for the Hamiltonian and the local operators; we call these required inputs the CFT data. After describing the theoretical foundations of the approach, we give details of an efficient algorithmic implementation. We illustrate the use of the algorithm by presenting recent examples of its applications, which range from evaluating static quantities such as spectra and operator matrix elements to non-equilibrium time evolution. 

Besides presenting the improved algorithm, a further goal of this work is to make the powerful method of TCSA more accessible for researchers working in low-dimensional quantum field theory. With this aim in view, we make available a MATLAB implementation \cite{scripts} of the algorithm, and also include two appendices which describe the computation of CFT data for two important classes of conformal field theories: minimal models and free boson CFTs. Note that the algorithm reported here and the code package \cite{scripts} can be applied to a much larger class of perturbed CFTs, requiring only the construction of the CFT data needed as input.

The CFTCSA algorithm has already been successfully applied to problems such as investigating the Kibble-Zurek scaling \cite{2020ScPP....9...55H}, time evolution after an inhomogeneous quench in the sine-Gordon field theory \cite{2021arXiv210906869H}, form factors of order/disorder fields \cite{2021arXiv210909767C} and also confinement of kinks \cite{2021arXiv211105360L} in perturbations of the tricritical Ising model. Finally, we note that independently of our work a similar approach was developed and applied to study the sinh-Gordon quantum field theory on a truncated free boson Hilbert space as reported in \cite{2021JHEP...01..014K}.

\section{Chirally Factorised Truncated Conformal Space Approach}\label{CFTCSAIdea}

\subsection{Off-critical deformations of conformal field theories}\label{subsec:PCFT}
In order to apply TCSA, the 1+1-dimensional quantum field theory under investigation is specified as the off-critical deformation of a conformal field theory (CFT) \cite{1984NuPhB.241..333B} by a relevant operator, with the Hamiltonian\footnote{Here the Hamiltonian is written in the Schr{\"o}dinger picture, or equivalently, taken at the fixed time $t=0$.}
\begin{equation}
    H=H_\mathrm{CFT}+\lambda\int_0^L \text{d}x \mathcal{V}(x) 
    \label{eq:QFTHam}
\end{equation}
where $H_\mathrm{CFT}$ is the CFT Hamiltonian, $\mathcal{V}(x)$ is the relevant operator and $\lambda$ is a coupling with positive mass dimension. It is possible to add multiple relevant fields by a trivial extension of the method described below. Space is taken to be a circle of finite length $L$: $x\equiv x+L$, corresponding to periodic boundary conditions. Under these circumstances, the conformal field theory has a symmetry algebra $\mathcal{A}\otimes\bar{\mathcal{A}}$ consisting of the tensor product of a left and a right \emph{chiral} factor. The Hilbert space of the theory can be decomposed according to irreducible representations of the symmetry algebra:
\begin{equation}
 \mathcal{H}_\mathrm{CFT}=\bigoplus_{\Phi} \mathcal{W}_\Phi
 \label{eq:HilbertCFT}
\end{equation}
with each subspace labelled by an irreducible representation of the algebra $\mathcal{A}\otimes\bar{\mathcal{A}}$ which are in one-to-one correspondence with so-called \emph{primary fields} $\Phi$. These fields transform in a product of irreducible representations $\mathcal{R}(\Phi)\otimes\bar{\mathcal{R}}(\Phi)$, and their behaviour under scaling transformations is specified by \emph{conformal weights} $(h_\Phi,\bar{h}_\Phi)$. To apply the TCSA method, it must be assumed that there is either a finite number or at most a countable infinity of these fields with a discrete spectrum of conformal weights bounded from below\footnote{CFTs with a finite number of primary fields are called \emph{rational} and their class includes numerous models relevant for field theory description of many-body systems, and statistical physics.}. We also assume that the Hamiltonian \eqref{eq:QFTHam} is translationally invariant, so the conformal weights of the perturbing field $\mathcal{V}$ satisfy $h_\mathcal{V}=\bar{h}_\mathcal{V}$.

To exploit the full power of conformal symmetry, it is useful to continue analytically to imaginary time $\tau=-it$ introducing complex coordinates $w=\tau-ix$ and $\bar{w}=\tau+ix$, and to map the space-time cylinder $(w,\bar{w})$ to the complex plane $(z,\bar{z})$ by the conformal transformation 
\begin{equation}
    z=\exp\frac{2\pi w}{L}\,,
\label{eq:confexpmao}
\end{equation}
which maps the cylinder to the complex plane parameterised by the dimensionless complex coordinates. Under such a map, primary fields given on the space-time cylinder transform as \cite{1984NuPhB.241..333B}
\begin{equation}
\mathcal{O}^{\mathrm{cyl}}(w,\bar{w})=
\left(\frac{2\pi z}{L}\right)^{h_\mathcal{O}}\left(\frac{2\pi\bar{z}}{L}\right)^{\bar{h}_\mathcal{O}}
\mathcal{O}^{\mathrm{pl}}(z,\bar{z})\,.\label{MapPlaneCyl}
\end{equation}
where the superscripts $\mathrm{cyl}$ and $\mathrm{pl}$ correspond to operators given on the space-time cylinder and on the complex plane, respectively.

For simplicity we now assume that the two chiral symmetry algebras are identical: $\mathcal{A}\equiv \bar{\mathcal{A}}$, which is the case for most applications of the TCSA. All subsequent considerations can be extended to the case with different left and right chiral algebras, but it makes the notations considerably less transparent, so we restrict ourselves to the case which is of practical interest. The subspaces corresponding to the primary fields are then given by 
\begin{equation}
    \mathcal{W}_\Phi=V_{\mathcal{R}(\Phi)}\otimes{{V}}_{\bar{\mathcal{R}}(\Phi)}
\label{eq:chiral_factorisation}    
\end{equation}
with the chiral factors $V_{\mathcal{R}(\Phi)}$ carry the irreducible representation $\mathcal{R}(\Phi)$ of the algebra $\mathcal{A}$. Notice that there can be fields transforming under different left and right representations; theories for which all fields satisfy $\mathcal{R}(\Phi)=\bar{\mathcal{R}}(\Phi)$ are called diagonal; the prime examples are given by the $A$ series of modular invariant partition functions \cite{Cappelli:1986hf,1987CMaPh.113....1C}. Denoting the generators of the chiral scaling transformations by $L_0$ and $\bar{L}_0$ as usual, the chiral spaces decompose into their eigenspaces:
\begin{align}
   V_\mathcal{R}&=\bigoplus_{N=0}^\infty V_\mathcal{R}(N) \,:\quad
   L_0\ket{v}=\left(h_\mathcal{R}+N\right)\ket{v}\quad\forall\ket{v}\in V_\mathcal{R}(N)
   \nonumber\\
   {V}_{\bar{\mathcal{R}}}&=\bigoplus_{\bar{N}=0}^\infty {V}_{\bar{\mathcal{R}}}(\bar{N}) \,:\quad
   \bar{L}_0\ket{\bar{v}}=\left(\bar{h}_{\bar{\mathcal{R}}}+\bar{N}\right)\ket{\bar{v}}\quad\forall\ket{\bar{v}}\in{V}_{\bar{\mathcal{R}}}(\bar{N})\,.
   \label{eq:level_spaces}
\end{align}
The dimensions of the chiral subspaces at fixed levels are denoted by 
\begin{equation}
    d_\mathcal{R}(N)=\dim V_\mathcal{R}(N)\,.
\end{equation}
In the simplest case, the \emph{highest weight subspace}  
$V_{\mathcal{R}(\Phi)}(0)\otimes{V}_{\bar{\mathcal{R}}(\Phi)}(0)$ of $\mathcal{W}_\Phi$ is spanned by a single vector $\ket{\Phi}$ which is created from the CFT vacuum by the action of the primary field $\Phi$. In models with current algebra symmetries the primary fields are themselves multiplets, e.g. for SU(2) WZNW models corresponding to diagonal modular invariant partition functions  \cite{Cappelli:1986hf} the primary field $\Phi_j$ transforms in the $(2j+1)\times(2j+1)$ dimensional representation of an $SU(2)\times SU(2)$ zero level subalgebra of $\mathcal{A}\otimes\bar{\mathcal{A}}$. Vectors with non-zero level $N+\bar{N}$ are called \emph{descendant vectors}, with their chiral descendant \emph{levels} given by $(N,\bar{N})$. 

Using \eqref{eq:chiral_factorisation}, the Hilbert space takes the form
\begin{equation}
     \mathcal{H}_\mathrm{CFT}=\bigoplus_{\Phi,N,\bar{N}} V_{\mathcal{R}(\Phi)}(N)\otimes{{V}}_{\bar{\mathcal{R}}(\Phi)}(\bar{N})
\,.\label{eq:factorised_Hilbert_space}
\end{equation}
Accordingly, any local operator $\mathcal{O}(x)$ with left/right conformal weights $h_\mathcal{O}$ and $\bar{h}_\mathcal{O}$ can be decomposed as
\begin{equation}
    \mathcal{O}(0)=\left(\frac{2\pi}{L}\right)^{h_\mathcal{O}+\bar{h}_\mathcal{O}}
    \bigoplus_{\genfrac{}{}{0pt}{3}{\Phi,N,\bar{N},}{\Phi',N',\bar{N}'}}\mathcal{C}_{\Phi'\Phi}\left(\mathcal{O}\right)
    \mathcal{B}^{\mathcal{O}}(\mathcal{R}(\Phi'),N',\mathcal{R}(\Phi),N)
    \otimes
    \bar{\mathcal{B}}^{\mathcal{O}}(\bar{\mathcal{R}}(\Phi'),\bar{N}',\bar{\mathcal{R}}(\Phi),\bar{N})
\label{eq:op_action_factorised}
\end{equation}
where $\mathcal{B}^{\mathcal{O}}(\mathcal{R}',N',\mathcal{R},N)$ and
$\bar{\mathcal{B}}^{\mathcal{O}}(\bar{\mathcal{R}}',\bar{N}',\bar{\mathcal{R}},\bar{N})$ are the \emph{chiral  three-point matrices} (a.k.a. chiral vertex operators) which only depend on the representations of the chiral algebra involved in the particular block, and $\mathcal{C}_{\Phi'\Phi}(\mathcal{O})$ are the \emph{operator product structure constants} of the CFT.

\subsection{The ``curse of dimensionality'' and the idea of CFTCSA}

In TCSA, the Hilbert space is truncated at some conformal weight i.e. imposing an upper limit of the eigenvalues of $L_0+\bar{L}_0$, making the space finite dimensional. The challenge is that the number of states grows very fast with the descendant level. In fact, the density of states has an exponential asymptotic behaviour \cite{Cardy:1986ie}
\begin{equation}
    \sim \exp\left\{ 2\pi\sqrt{\frac{c(h+N-c/24)}{6}}+2\pi\sqrt{\frac{c(\bar{h}+\bar{N}-c/24)}{6}}\right\}\,,
\end{equation}
where $c$ is the central charge of the CFT. As a result, while truncations with only a few dozen states work surprisingly well for simple cases \cite{1990IJMPA...5.3221Y,1991IJMPA...6.4557Y}, they quickly become insufficient for other models where the central charge is higher, or the convergence with the cut-off is slower. While this can be helped by renormalisation group methods as mentioned in the introduction, the perturbative running of the coupling is only valid for sufficiently high cut-offs, which leaves one with the challenge of accommodating very large Hilbert spaces, and subsequently very large operator matrices in computer memory. Note that the full space occupied by the operator matrices grows with the square $D^2$ of the dimension $D$ of the truncated Hilbert space. In many cases, these matrices are not even sparse enough to gain substantially from the exploitation of their structure.

However, the chiral factorisation property \eqref{eq:chiral_factorisation} implies that the size of the CFT data considered as a function of the cut-off, only grows with the square root of the size of the complete (nonchiral) operator matrices, i.e. effectively proportional to the dimension $D$ of the truncated Hilbert space instead of $D^2$. Therefore, an algorithm working directly in terms of the chiral data improves efficiency by quite a large margin, despite still being ultimately subject to the same exponential growth\footnote{We remark that further improvements (albeit more moderate than from chiral factorisation) can be achieved by exploiting specific structural properties of the operator matrices, as mentioned at the end of Appendix \ref{app:vertexopmatrix}.}.

The idea of the Chirally Factorised TCSA is to set up the numerical algorithm exploiting \eqref{eq:chiral_factorisation}, and use it to reduce the size of the CFT input data needed to set up the TCSA. 

The CFT data required as input to the CFTCSA algorithm are the following:
\begin{itemize}
    \item The central charge $c$ of the CFT, which is a parameter of the chiral algebra $\mathcal{A}$. More precisely, it is a parameter of its Virasoro sub-algebra $Vir$ which is always present as a consequence of the conformal symmetry.
    \item The decomposition \eqref{eq:HilbertCFT} of the CFT Hilbert space, which specifies the full spectrum of the conformal field theory and is severely constrained by the requirement of modular invariance \cite{1987CMaPh.113....1C}.
    \item{The matrix elements of the perturbing field with highest weight vectors}
    \begin{equation}
        \mathcal{C}_{\Phi'\Phi}\left(\mathcal{V}\right)=\bra{\Phi'}\mathcal{V}(0)\ket{\Phi}
    \end{equation}
    which are examples of the operator product structure constants of the CFT. Matrix elements of the perturbing field with descendant vectors can then be determined using the symmetry algebra $\mathcal{A}\otimes\bar{\mathcal{A}}$ to construct the relevant chiral three-point matrices (c.f. Appendix \ref{sec:Virasoro} for the Virasoro case). To evaluate matrix elements of any other local operator $\mathcal{O}$, it is also necessary to include the values of the structure constants
    \begin{equation}
        \mathcal{C}_{\Phi'\Phi}\left(\mathcal{O}\right)=\bra{\Phi'}\mathcal{O}(0)\ket{\Phi}
    \end{equation}
    \item{The matrix elements of the chiral three-point matrices 
    \begin{equation}
    \mathcal{B}^{\mathcal{O}}(\mathcal{R}',N',\mathcal{R},N)_{\alpha'\alpha}
    \quad \textrm{and}\quad 
    \bar{\mathcal{B}}^{\mathcal{O}}(\mathcal{R}',\bar{N}',\mathcal{R},\bar{N})_{\bar{\alpha}'\bar{\alpha}}
    \end{equation}
    at least for the perturbing operator $\mathcal{O}=\mathcal{V}$, and also for other operators if necessary for the particular physical problem considered.}
\end{itemize}
With the exception of the chiral three-point matrices, the size of all the other data is independent of the cut-off and essentially negligible in comparison: the real gain of the CFTCSA algorithm comes from encoding the operator matrices in terms of the chiral three-point matrices.

The Hamiltonian of the conformal field theory is given by
\begin{equation}
    H_\mathrm{CFT}=\frac{2\pi}{L}\left(L_0+\bar{L}_0-\frac{c}{12}\right)
    \label{eq:HCFT}
\end{equation}
and it is automatically diagonal in the basis specified by \eqref{eq:level_spaces}. Following \eqref{eq:op_action_factorised}, the matrix elements of a local primary field $\mathcal{O}(x)$ between two basis vectors
\begin{align}
    &\ket{w}=\ket{\mathcal{R}(\Phi),N,\alpha}\otimes\ket{\bar{\mathcal{R}}(\Phi),\bar{N},\bar{\alpha}}\,\in\, V_{\mathcal{R}(\Phi)}(N)\otimes V_{\bar{\mathcal{R}}(\Phi)}(\bar{N})\nonumber\\
    &\ket{w'}=\ket{\mathcal{R}(\Phi'),N',\alpha'}\otimes\ket{\bar{\mathcal{R}}(\Phi'),\bar{N}',\bar{\alpha}'}\,\in\, V_{\mathcal{R}(\Phi')}(N')\otimes V_{\bar{\mathcal{R}}(\Phi')}(\bar{N'})
\end{align}
can be written (using translational invariance) as follows:
\begin{align}
    \bra{w'}\mathcal{O}(x)\ket{w}=&\left(\frac{2\pi}{L}\right)^{h_\mathcal{O}+\bar{h}_\mathcal{O}}\mathcal{C}_{\Phi'\Phi}\left(\mathcal{O}\right)
    \mathcal{B}^{\mathcal{O}}(\mathcal{R}(\Phi'),N',\mathcal{R}(\Phi),N)_{\alpha'\alpha}
    \bar{\mathcal{B}}^{\mathcal{O}}(\bar{\mathcal{R}}(\Phi'),\bar{N}',\bar{\mathcal{R}}(\Phi),\bar{N})_{\bar{\alpha}'\bar{\alpha}}
    \nonumber\\
    &\cdot\exp\left(i\frac{2\pi}{L}
    \left[
    \left(h_{\Phi'}+N'-\bar{h}_{\Phi'}-\bar{N}'\right)
    -\left(h_{\mathcal{O}}-\bar{h}_{\mathcal{O}}+h_{\Phi}+N-\bar{h}_{\Phi}-\bar{N}\right)
    \right]x\right)\,,
    \label{eq:Vmatrix}
\end{align}
where we assumed that the indices $\alpha=1,\dots,d_\mathcal{R}(N)$ run over an orthonormal basis of the level subspaces $V_\mathcal{R}(N)$. To fix normalisation conventions for the structure constants, it is also necessary to specify the normalisation of the chiral three-point matrices $\mathcal{B}^\mathcal{O}$ and $\bar{\mathcal{B}}^\mathcal{O}$ for primary operators $\mathcal{O}$. For the case when the zero-level subspaces are one-dimensional, they are normalised as
\begin{equation}
    \mathcal{B}^{\mathcal{O}}(\mathcal{R}',0,\mathcal{R},0)=1\quad ,\quad 
    \bar{\mathcal{B}}^{\mathcal{O}}(\bar{\mathcal{R}}',0,\bar{\mathcal{R}},0)=1\, .
\label{eq:Bnormalisation}
\end{equation}
For the case when the zero-level subspaces are themselves multiplets under some symmetry group, the zero-level parts of $\mathcal{B}$ and $\bar{\mathcal{B}}$ are fixed in terms of the appropriate Clebsh-Gordan coefficients \cite{2015NuPhB.899..547K}. Given the zero-level parts, the descendant parts of $\mathcal{B}$ and $\bar{\mathcal{B}}$ are fully determined by the symmetry algebra $\mathcal{A}\otimes\bar{\mathcal{A}}$. In the following we assume these matrices have been constructed by appropriate CFT calculations. Note that strictly speaking, for the algorithm described in this work the details of those construction are not relevant, as long as its results are available to provide the necessary inputs for the TCSA calculations. Nevertheless, we describe their construction for two different chiral algebras in Appendices \ref{sec:Virasoro} and \ref{sec:boson}.

\subsection{Vectors, inner products and operator matrix elements}

A general vector in the conformal Hilbert space \eqref{eq:HilbertCFT} can be written as  
\begin{equation}
|\Psi\rangle=\sum_{\Phi,N,\bar{N}}\sum_{\alpha=1}^{d_{\mathcal{R}(\Phi)}(N)}\sum_{\bar{\alpha}=1}^{d_{\bar{\mathcal{R}}(\Phi)}(\bar{N})}K_{\Psi}(\Phi,N,\bar{N})_{\alpha\bar{\alpha}}\ket{\mathcal{R}(\Phi),N,\alpha}\otimes\ket{\bar{\mathcal{R}}(\Phi),\bar{N},\bar{\alpha}}\,,
\label{eq:genvector}
\end{equation}
where $K_{\Psi}(\Phi,N,\bar{N})_{\alpha\bar{\alpha}}$ are complex vector coefficients represented as two-index tensors. Inner products can then be computed using 
\begin{equation}
\langle\Psi_1|\Psi_2\rangle =
\sum_{\Phi,N,\bar{N}}
\sum_{\alpha,\bar{\alpha}}
K_{\Psi_1}^{*}(\Phi,N,\bar{N})_{\alpha\bar{\alpha}}
K_{\Psi_2}(\Phi,N,\bar{N})_{\alpha\bar{\alpha}}
=\sum_{\Phi,N,\bar{N}}\text{Tr} \left\{
K_{\Psi_1}(\Phi,N,\bar{N})^{\dagger}K_{\Psi_2}(\Phi,N,\bar{N}) \right\}
\label{eq:innerproduct}
\end{equation}
Using \eqref{eq:Vmatrix}, the matrix elements of a local scaling field $\mathcal{O}(x)$ can be written  as
\begin{align}
&\langle\Psi_1|\mathcal{O}(x)|\Psi_2\rangle=
\nonumber\\
&\left(\frac{2\pi}{L}\right)^{h_\mathcal{O}+\bar{h}_\mathcal{O}}\sum_{\Phi_1,\Phi_2}\Bigg\{
\mathcal{C}_{\Phi_1\Phi_2}(\mathcal{O})
\nonumber\\
&\sum_{{N_1,\bar{N}_1}}
\sum_{{N_2,\bar{N}_2}}
\text{Tr}\left\{
K_{\Psi_1}(\Phi_1,N_1,\bar{N}_1)^\dagger
\mathcal{B}^\mathcal{O}(\mathcal{R}(\Phi_1),N_1,\mathcal{R}(\Phi_2),N_2)
K_{\Psi_2}(\Phi_2,N_2,\bar{N}_2)
\bar{\mathcal{B}}^\mathcal{O}(\bar{\mathcal{R}}(\Phi_1),\bar{N}_1,\bar{\mathcal{R}}(\Phi_2),\bar{N}_2)^T \right\}
\nonumber\\
&\qquad\cdot\exp\left(i\frac{2\pi}{L}
    \left[
    \left(h_{\Phi_1}+N_1-\bar{h}_{\Phi_1}-\bar{N}_1\right)
    -\left(h_{\mathcal{O}}-\bar{h}_{\mathcal{O}}+h_{\Phi_2}+N_2-\bar{h}_{\Phi_2}-\bar{N_2}\right)
    \right]x\right)\Bigg\}\,.
\label{eq:localoperator}
\end{align}

\subsection{Representing the perturbing operator}\label{subsec:pert_operator_repr}
Due to the assumption of translational invariance, the matrix elements of the perturbation 
\begin{equation}
    H_\mathrm{pert}=\int_0^L \text{d}x \mathcal{V}(x) 
    \label{eq:pertHam}
\end{equation}
can be simplified with respect to the general formula \eqref{eq:localoperator}. Namely, the condition $h_\mathcal{V}=\bar{h}_\mathcal{V}$ and the spatial integral leads to a superselection rule according to the momentum $P$, which is given by
\begin{equation}
    P\left(\ket{\Phi,N,\alpha}\otimes\ket{\Phi,\bar{N},\bar{\alpha}}\right)
    =\frac{2 \pi}{L}\left(s_\Phi+N-\bar{N}\right)
    \left(\ket{\Phi,N,\alpha}\otimes\ket{\Phi,\bar{N},\bar{\alpha}}\right)
    \,,
\end{equation}
where $s_\Phi=h_\Phi-\bar{h}_\Phi$ is the conformal spin of the primary field $\Phi$. For a state with momentum $2\pi s/L$, its components in the representation \eqref{eq:genvector} satisfy the selection rule
\begin{equation}
K_{\Psi}(\Phi,N,\bar{N})=0\quad\text{for}\quad s_\Phi+N-\bar{N}\neq s\,,
\end{equation}
and taking momentum eigenstates $\ket{\Psi_{1,2}}$ with momentum eigenvalues given by
\begin{equation}
    P\ket{\Psi_{1,2}}=\frac{2\pi}{L}s(\Psi_{1,2})\ket{\Psi_{1,2}}
\end{equation}
the spatial integral of the matrix element \eqref{eq:localoperator} can be rewritten as
\begin{align}
&\langle\Psi_1|H_\textrm{pert}|\Psi_2\rangle=
\nonumber\\
&\left(\frac{2\pi}{L}\right)^{2h_\mathcal{V}}
\sum_{\Phi_1,\Phi_2}
\mathcal{C}_{\Phi_1\Phi_2}(\mathcal{O})
\sum_{{N_1,\bar{N}_1}}
\sum_{{N_2,\bar{N}_2}}
\Bigg\{L \delta_{s(\Psi_1),s(\Psi_2)}
\nonumber\\
&\cdot\text{Tr}\left\{
K_{\Psi_1}(\Phi_1,N_1,\bar{N}_1)^\dagger
\mathcal{B}^\mathcal{V}(\mathcal{R}(\Phi_1),N_1,\mathcal{R}(\Phi_2),N_2)
K_{\Psi_2}(\Phi_2,N_2,\bar{N}_2)
\bar{\mathcal{B}}^\mathcal{V}(\bar{\mathcal{R}}(\Phi_1),\bar{N}_1,\bar{\mathcal{R}}(\Phi_2),\bar{N}_2)^T  \right\}\Bigg\}
\,.
\label{eq:Hpertmatrix}
\end{align}
As a result, the action of the perturbing operator on a momentum eigenstate $\ket{\Psi}$
\begin{equation}
    \ket{\Psi'}=H_\mathrm{pert}\ket{\Psi}
\end{equation}
gives a state $\ket{\Psi'}$ with the same momentum, and can be written in terms of the vector components as
\begin{align}
&K_{\Psi'}(\Phi',N',\bar{N}')_{\alpha'\bar{\alpha}'}=
\nonumber\\
&
\left(\frac{2\pi}{L}\right)^{2h_\mathcal{V}}
\sum_{{\Phi,N,\bar{N}}}L \delta_{s',s}
\sum_{\genfrac{}{}{0pt}{3}{\alpha,\bar{\alpha}}{\alpha',\bar{\alpha}'} }
\mathcal{C}_{\Phi'\Phi}(\mathcal{O})
\mathcal{B}^\mathcal{V}(\mathcal{R}(\Phi'),N',\mathcal{R}(\Phi),N)_{\alpha',\alpha}
\bar{\mathcal{B}}^\mathcal{V}(\bar{\mathcal{R}}(\Phi'),\bar{N}',\bar{\mathcal{R}}(\Phi),\bar{N})_{\bar{\alpha'},\bar{\alpha}} 
K_{\Psi}(\Phi,N,\bar{N})_{\alpha\bar{\alpha}}
\nonumber\\
&\textrm{where}\quad s=s_\Phi+N-\bar{N} \quad\text{and}\quad s'=s_{\Phi'}+N'-\bar{N}'\,,
\end{align}
or alternatively in a compact matrix notation as
\begin{align}
&K_{\Psi'}(\Phi',N',\bar{N}')=
\nonumber\\
&
\left(\frac{2\pi}{L}\right)^{2h_\mathcal{V}}
\sum_{\Phi,N,\bar{N}}
L \delta_{s',s}\,
\mathcal{C}_{\Phi'\Phi}(\mathcal{O})
\mathcal{B}^\mathcal{V}(\mathcal{R}(\Phi'),N',\mathcal{R}(\Phi),N)
 K_{\Psi}(\Phi,N,\bar{N})
\bar{\mathcal{B}}^\mathcal{V}(\bar{\mathcal{R}}(\Phi'),\bar{N}',\bar{\mathcal{R}}(\Phi),\bar{N})^T
\label{eq:Hpertaction}
\end{align}
We emphasise that due to the conservation of momentum, for the action of a translationally invariant Hamiltonian the Hilbert space can be limited to states with a fixed value $s$, corresponding to a subspace with a given total momentum $2\pi s/L$. However, matrix elements \eqref{eq:localoperator} of local operators between vectors in different momentum sectors are in general nonzero.

\section{Implementation}
\label{sec:implementation}

\subsection{Describing the Hilbert space}
\label{sec:implementation:H-ChDescriptors}

The key idea behind our improved numerical approach is to exploit the chiral factorisation \eqref{eq:chiral_factorisation} to optimise memory usage and performance. In virtue of \eqref{eq:chiral_factorisation}, the basis states can be considered expressed as pairs of basis states from the chiral subspaces $V_R$. This chiral structure can be encoded in \emph{descriptors} which are matrices describing the Hilbert space and local operators in terms of the chiral building blocks.

The \emph{Chiral Descriptor} summarises the basic information about the chiral subspaces  $V_\mathcal{R}(N)$. Since for every primary field these spaces are specified by giving the representations $\mathcal{R}_\Phi$ and $\bar{\mathcal{R}}_\Phi$, the independent information needed is to list at each level $N$ the subspaces $V_\mathcal{R}(N)$ of all the chiral representations $\mathcal{R}$. These can be ordered by increasing chiral conformal weight ($L_0$ eigenvalue) $h_\mathcal{R}+N$, and the multiindex $(\mathcal{R},N)$ replaced by their position $n$  in this list (in case of subspaces of equal weight, their order can be chosen in an arbitrary way); however, depending on the problem, other orderings may be more convenient. The CFT data constructed for the purpose of the TCSA computations must contain the basis of the chiral level subspaces to some upper limit $h(\mathcal{R},N)<h_\text{max}$ which is chosen sufficiently high to contain all states that occur in the numerical computations. The Chiral Descriptor is then a two-column matrix listing the conformal weights $h_n$ and dimensions $d_n$ of the chiral level subspaces $V_\mathcal{R}(N)$:  
\begin{equation}
D_\mathrm{Ch} = 
\begin{pmatrix}
h_{1} & d_1 \\
h_{2} & d_2 \\
h_{3} & d_3 \\
\vdots & \vdots
\end{pmatrix}\,,
\label{eq:chiral_descriptor}
\end{equation}
with $h_1\leq h_2\leq h_3\leq \dots$.

The primary fields $\Phi$ appearing in the decomposition \eqref{eq:HilbertCFT} can be enumerated in some particular order as 
\begin{equation}
\Phi_M\, ,\, M=1,\dots l_\textrm{Pr}\,.
    \label{eq:primary_list}
\end{equation}
For CFTs with a finite number of primary fields (so-called rational CFTs), $l_\textrm{Pr}$ is a fixed number, but for those with infinitely many primaries (such as the free massless boson) this list must be terminated so that every primary field $\Phi$ appearing as a subspace $\mathcal{W}_\Phi$ in the truncated Hilbert space is included. Typically, this list is ordered by increasing scaling dimension, with the identity operator coming first, although depending on the problem other orderings may be more convenient.

For example, in the Ising CFT with central charge $c=1/2$ the primaries are the identity $\Phi_1=\mathbb{I}$, the magnetisation $\Phi_2=\sigma$ and the energy operator $\Phi_3=\epsilon$, with conformal dimensions $h_1=\bar{h}_1=0$, $h_2=\bar{h}_2=1/16$ and $h_3=\bar{h}_3=1/2$.  

The Hilbert space \eqref{eq:HilbertCFT} is specified by a \emph{Hilbert Space Descriptor} which prescribes how to sew together the left- and right-handed chiral spaces. It is a three-column matrix with the indices of the left- and right-handed subspaces in its first and second columns, respectively, while the third column contains the index of the primary field corresponding to the subspace. The indices refer to the rows of the Chiral Descriptor and therefore specify the conformal weight and the dimensionality of the subspaces in the product space.

For the simplest, but also most frequently considered example of the zero-momentum subspace in a CFT with diagonal partition function, the Hilbert Space Descriptor has a particularly simple form. E.g., for the Ising model, the first four rows correspond to $V_{\mathcal{R}(\mathbb{I})}(0)\otimes V_{\mathcal{R}(\mathbb{I})}(0)$, $V_{\mathcal{R}(\sigma)}(0)\otimes V_{\mathcal{R}(\sigma)}(0)$, $V_{\mathcal{R}(\epsilon)}(0)\otimes V_{\mathcal{R}(\epsilon)}(0)$ and $V_{\mathcal{R}(\sigma)}(1)\otimes V_{\mathcal{R}(\sigma)}(1)$, and the Hilbert Space Descriptor takes the form
\begin{equation}
D_\mathrm{H} = \begin{pmatrix}1 & 1 & 1\\
2 & 2 & 2\\
3 & 3 & 3\\
4 & 4 & 2\\
\vdots & \vdots & \vdots
\end{pmatrix}\,.
\label{eq:diagonal_zeromomentum}
\end{equation}
where the first and second column entries refer to the corresponding Chiral Descriptor, the first four lines of which describe the chiral level subspaces $V_{\mathcal{R}(\mathbb{I})}(0)$, $V_{\mathcal{R}(\sigma)}(0)$, $V_{\mathcal{R}(\epsilon)}(0)$ and $V_{\mathcal{R}(\sigma)}(1)$: 
\begin{equation}
D_\mathrm{Ch} = 
\begin{pmatrix}
0 & 1\\
1/16 & 1\\
1/2 & 1\\
17/16 & 1\\
\vdots & \vdots
\end{pmatrix}\,,
\label{eq:chiral_descriptor_Ising}
\end{equation}

If the $m$th row of $D_\mathrm{H}$ has $j,k$ as its first two elements, then the corresponding subspace has total conformal weight $D_\mathrm{Ch}(j,1) + D_\mathrm{Ch}(k,1)$, conformal spin $D_\mathrm{Ch}(j,1) - D_\mathrm{Ch}(k,1)$, and dimension $D_\mathrm{Ch}(j,2) \times D_\mathrm{Ch}(k,2)$. As the TCSA introduces an upper limit on the total conformal weight, the number of included subspaces are limited to a finite value $l_\mathrm{H}$. 

The Hilbert Space Descriptor provides the recipe to express the general state vectors \eqref{eq:genvector} as
\begin{equation}
    \ket{\Psi}=K^\Psi(m)_{\alpha\beta} \ket{m,\alpha,\beta}\,,
\end{equation}
which are most conveniently handled as lists of matrices $K(m)$, $m=1,\dots,l_\mathrm{H}$ with sizes dictated by $D_\mathrm{H}$ and $D_\mathrm{Ch}$.  From \eqref{eq:innerproduct}, the inner products of two vectors $\ket{\Psi_1}$ and $\ket{\Psi_2}$ is given in by
\begin{equation}
\label{eq:matelem_intloc}
    \bra{\Psi_1}\ket{\Psi_2} = \sum_{m = 1}^{l_\mathrm{H}} \Tr{K^{\Psi_1}(m)^\dagger K^{\Psi_2}(m)}\,.
\end{equation}

\subsection{Matrix elements of local operators}
\label{ssec:matrixelem}

The action of any scaling operator on the Hilbert space is specified by the chiral three-point matrices $\mathcal{B}^{\mathcal{O}}(\mathcal{R},N,\mathcal{R}',N')$ introduced in \eqref{eq:Vmatrix}. Using the Chiral Descriptor number $n$ for the level subspace labelled by $(\mathcal{R},N)$, the CFT data must contain a list of all necessary chiral three-point matrices $\mathcal{B}^{\mathcal{O}}(n,n')$ and $\bar{\mathcal{B}}^{\mathcal{O}}(n,n')$ for all operators $\mathcal{O}$ that occur in the computation, conveniently arranged as a list of matrices $\underrightarrow{\mathcal{B}}$ which we call the \emph{Operator List}.

The action of a scaling operator $\mathcal{O}$ is then (partially) specified using two (left/right)
\emph{Operator Descriptors} $D^{\mathcal{O},L}_{\text{Op}}$ and $D^{\mathcal{O},R}_{\text{Op}}$. A computationally efficient representation of the operator descriptor is a square matrix filled with integers (illustrated for the left): 
\begin{equation}
    D^{\mathcal{O},L}_{\text{Op}}=\begin{pmatrix}0 & 0 & 0 & 0 & 1 & 0 & 0 & \cdots\\
0 & 0 & 0 & 0 & 0 & 2 & 0&\cdots\\
0 & 0 & 0 & 3 & 0 & 0 & 4&\cdots\\
\vdots &  &  & \vdots &  &  &  & \ddots
\end{pmatrix}\,,
\end{equation}
and a similar descriptor must be specified for the right chiral part. We call these quantities \emph{Operator Descriptor Matrices}, which have the following meaning. The rows and columns correspond to the rows of the Hilbert Space Descriptor, listing the chiral decomposition of the level subspaces of the full Hilbert space. Assuming that the $n$th row of the Hilbert space descriptor corresponds to the subspace $V_\mathcal{R}(N)\otimes V_{\mathcal{\bar{R}}}(\bar{N})$, while the $n'$th row corresponds to the subspace $V_{\mathcal{R}'}(N')\otimes  V_{\bar{\mathcal{R}}'}(\bar{N}')$, the element $D^{\mathcal{O},L}_{\text{Op}}(n,n')$ gives the position of the left chiral three-point matrix  $\mathcal{B}^{\mathcal{O}}(\mathcal{R},N,\mathcal{R}',N')$ in the Operator List $\underrightarrow{\mathcal{B}}$, or zero if the chiral three-point matrix vanishes, i.e., we define $\underrightarrow{\mathcal{B}}(0)=0$ by convention. Similarly, the element $D^{\mathcal{O},R}_{\text{Op}}(n,n')$ specifies the position of the right chiral three-point matrix  $\bar{\mathcal{B}}^{\mathcal{O}}(\bar{\mathcal{R}},\bar{N},\bar{\mathcal{R}}',\bar{N}')$ in the Operator List $\underrightarrow{\mathcal{B}}$. Note that the left/right blocks are identical for operators with $\mathcal{R}\left(\mathcal{O}\right)=\bar{\mathcal{R}}\left(\mathcal{O}\right)$, i.e. transforming in the same representation both on the left and the right. In such a case the two descriptors are identical and can be given by the same matrix: $D^{\mathcal{O},L}_{\text{Op}}=D^{\mathcal{O},R}_{\text{Op}}=D_{\text{Op}}^{\mathcal{O}}$. The dimensions of the Operator Descriptor Matrices are  $l_{\text{H}}\cross l_{\text{H}}$.

In some cases it is more convenient to have separate  Operator Lists $\underrightarrow{\mathcal{B^O}}$ instead of the single unified list  $\underrightarrow{\mathcal{B}}$. This alters the meaning of the elements of $D^{\mathcal{O},L}_{\text{Op}}$ and $D^{\mathcal{O},R}_{\text{Op}}$, as they now index the Operator List of $\mathcal{O}$. This is especially useful if only some selected operators are relevant to a given physical problem, which is in fact the case for the examples in Section \ref{sec:appications}. 

In addition, in certain problems alternative choices can be more optimal than the generic Operator Descriptor Matrices and Operator Lists introduced above. An example of an alternative realisation which is optimised for the case of creation/annihilation operators in the free boson CFT is described in Appendix \ref{AppB4}.

To further facilitate implementation, the definition of the operator algebra structure constants $\mathcal{C}_{\Phi\Phi'}(\mathcal{O})$ is encoded as follows:
\begin{equation}
    \mathcal{C}^{\mathcal{O}}(M,M')=\mathcal{C}_{\Phi\Phi'}(\mathcal{O})
\end{equation}
where $\Phi$ (resp. $\Phi'$) is the primary field corresponding to the module $\mathcal{W}_\Phi$ and $\mathcal{C}^{\mathcal{O}}(M,M')$ is a rewriting of $\mathcal{C}_{\Phi\Phi'}(\mathcal{O})$ in which the primary fields are indexed according to \eqref{eq:primary_list} by integers $M$ and $M'>0$, in accordance with the 3rd row of the Hilbert Space Descriptor.  In the particular programming implementation the structure constants $\mathcal{C}^{\mathcal{O}}(M,M')$ are stored in a matrix form, which we call the \emph{Structure Constant Matrix}. The non-zero elements at position $(M,M')$ are the actual structure constants connecting the conformal family of the operator $\mathcal{O}$ under consideration and the conformal families associated with $M$th and $M'$th primaries, whose left and right chiral parts, similarly to $\mathcal{O}$ itself, can be different in a generic CFT. Given these considerations, the dimension of this matrix is $l_{\text{Pr}}\cross l_{\text{Pr}}$, where $l_{\text{Pr}}$ denotes the number of primary fields in the theory\footnote{On the number of primaries $l_{\text{Pr}}$ see the discussion after Eq. \eqref{eq:primary_list}.}. 

Finally, we would like to mention that the descriptors $D^{\mathcal{O},L/R}_{\text{Op}}$ and the Structure Constant Matrix $\mathcal{C}^{\mathcal{O}}(M,M')$ may include some redundancy due to superselection rules corresponding signalled by vanishing structure constants. Although this redundancy can be removed, we do not address this issue here for the sake of transparency and simplicity.

With these notations, the matrix elements of local operators \eqref{eq:localoperator} can eventually be computed as
\begin{equation}
\begin{split}
    \bra{\Psi_1}\mathcal{O}(x)\ket{\Psi_2} =\left(\frac{2\pi}{L}\right)^{h_\mathcal{O}+\bar{h}_\mathcal{O}}\sum_{m,m'=1}^{l_\mathrm{H}} \Bigg\{ \mathrm{Tr}
    \Big\{
    &\mathcal{C}^{\mathcal{O}}(D_\mathrm{H}(m,3),D_\mathrm{H}(m',3)) 
    K^{\Psi_1}(m)^{\dagger} \\
    &\underrightarrow{\mathcal{B}}\left(D^{\mathcal{O},L}_{\text{Op}}\left(m,m'\right)\right) K^{\Psi_2}(m')
    \underrightarrow{\mathcal{B}}\left(D^{\mathcal{O},R}_{\text{Op}}\left(m,m'\right)\right)^{T}
    \Big\}\times \\[0.5em]
    &\exp\left(i\frac{2\pi}{L}
    \left[s-s'-h_{\mathcal{O}}+\bar{h}_{\mathcal{O}}\right]x\right)
    \Bigg\}\,,
\end{split}
\label{eq:matelem_general}
\end{equation}
while the action \eqref{eq:Hpertaction} of an integrated spin-$0$ field 
\begin{equation}
    \ket{\Psi'}=\left(\int_0^{L}\text{d}x \mathcal{V}(x)\right)\ket{\Psi}
\end{equation}
can be computed as 
\begin{align}
K_{\Psi'}(m')=\left(\frac{2\pi}{L}\right)^{2h_\mathcal{V}}
\sum_{m=1}^{l_\mathrm{H}} &
L \delta_{s',s}\,
\mathcal{C}^{\mathcal{V}}(D_\mathrm{H}(m',3),D_\mathrm{H}(m,3))\times
\nonumber\\
&\times \underrightarrow{\mathcal{B}}\left(D^{\mathcal{V},L}_{\text{Op}}\left(m',m\right)\right) K_{\Psi}(m)
 \underrightarrow{\mathcal{B}}\left(D^{\mathcal{V},R}_{\text{Op}}\left(m',m\right)\right)^T\,,
 \label{eq:localOpAction}
\end{align}
where
\begin{equation}
    s=D_\mathrm{Ch}(D_\mathrm{H}(m,1),1)-D_\mathrm{Ch}(D_\mathrm{H}(m,2),1)\,,
\end{equation}
and similarly for $s'.$
We include a particular realisation of the above algorithms implemented using MATLAB \cite{MATLAB} as an ancillary file package \cite{scripts}, which also includes some examples of physical applications. 

In the implemented package, Operator Descriptor Matrices are expected to be exported as a `.dat' file, and can be imported by MATLAB's \texttt{load()} function, which guarantees that the Operator Descriptor Matrices are of the form of a $l_{\text{H}}\times l_{\text{H}}$ MATLAB array, whose elements can be reached as \texttt{OpDescriptorMatrix(m,m')} where \texttt{OpDescriptorMatrix} denotes the equivalent of $D_{\text{Op}}^{\mathcal{O}}$ in MATLAB (for operators with $\mathcal{R}\left(\mathcal{O}\right)\neq\bar{\mathcal{R}}\left(\mathcal{O}\right)$ there are two different matrices separately for left/right). The structure constants, again assuming a `.dat' file format, can again be imported by MATLAB's \texttt{load()} function, guaranteeing that the Structure Constant Matrix is of the form of a $l_{\text{Pr}}\times l_{\text{Pr}}$ MATLAB array, whose elements can be reached as \texttt{StructConstMatrix(M,M')} where \texttt{StructConstMatrix} denotes the equivalent of $\mathcal{C}^{\mathcal{O}}(M,M') $ in MATLAB.

Below we present an actual yet simplified script, more precisely a function written in MATLAB to better demonstrate the above ideas via the particular example of the action of an operator on a vector. A similar function computing matrix elements of operators is then straightforward to obtain.  We would like to stress that this code has to be regarded as a ``vanilla'' version of our MATLAB function and is presented in order to better and transparently demonstrate the logic of the actual implementation. The more elaborate and ready-to-use version of this function can be found online in \cite{scripts}.

\lstset{language=MATLAB}
\begin{footnotesize}
\begin{lstlisting}
function NewVector=ApplyOperator(VectorInTensors,ChiralDescriptor,HDescriptor, ...
OpDescriptorMatrixL,OpDescriptorMatrixR,StructConstMatrix)
l=size(HDescriptor,1);
NewVectorInTensors={};
global OpList;
for i=1:l
    clearvars dummy
    dummy=0*VectorInTensors{i};
    for j=1:l
        if  OpDescriptorMatrixL(i,j)~=0 &&
            OpDescriptorMatrixR(i,j)~=0
            
            dummy=dummy+StructConstMatrix(HDescriptor(i,3),HDescriptor(j,3))*
                        ((OpList{OpDescriptorMatrixL(i,j)}
                    *VectorInTensors{j})  
                    *(OpList{OpDescriptorMatrixR(i,j)}.'));
        end
    end
    NewVectorInTensors{i}=dummy;
end
NewVector=NewVectorInTensors;
end
\end{lstlisting}
\end{footnotesize}

The Operator List $\underrightarrow{\mathcal{B}}$ is expected to be prepared as a list of matrices named \texttt{OpList} in a `.mat' file, which can imported into MATLAB can be achieved by either the function \texttt{ImportMatFiles} or the function \texttt{ImportMatFilesNoSparse}; the latter results in sparse matrices which help reduce the use of memory. The individual matrices in the list are referred to as  \texttt{OpList\{k\}}, and are conventional (sparse or nonsparse) matrices with their elements addressed as \texttt{OpList\{k\}(l,m)}. One technical aspect we would like to emphasise is that the variable \texttt{OpList} is not passed to the above function as an argument, as it is a large object common throughout all calculations, and instead it is defined as a global variable.

\subsection{Spectral problem}

Most applications of the CFTCSA involve the solution of the spectral problem, i.e., finding eigenvalues and eigenvectors of a perturbed CFT defined by a Hamiltonian of the form \eqref{eq:QFTHam}. The action of the conformal Hamiltonian on a state can be computed easily
\begin{equation}
    K^{H_\mathrm{CFT}\Psi}(n) = \frac{2\pi}{L}\left(D_\mathrm{Ch}(D_\mathrm{H}(n,1),1)+D_\mathrm{Ch}(D_\mathrm{H}(n,2),1)-\frac{c}{12}\right)K^\Psi(n)\,,
    \label{eq:confHamaction}
\end{equation}
where $L$ is the finite volume parameter and $c$ is the central charge of the fixed point CFT, while the action of the perturbing fields is expressed through Eq. \eqref{eq:localOpAction}. Combining the two yields the map from $\ket{\Psi}$ to $H\ket{\Psi}$ without explicitly calculating and storing the matrix elements of $H$. The spectral problem is then solved using iterative eigensolvers which can operate having access to the action of the matrix only, such as e.g. the built-in eigensolver of MATLAB \cite{MATLAB}.

\subsection{Time evolution}

\subsubsection{Chebyshev-Bessel method}\label{subsec:CBmethod}
The time evolved state $e^{-itH}|\Omega\rangle$ can be computed using the Chebyshev-Bessel method \cite{2016NuPhB.911..805R}, which only needs the action of the Hamiltonian $H$ on vectors.  To evaluate the time evolution of the system starting from quantum state $|\Omega\rangle$, the time evolution operator $e^{-itH}$ is expanded on the basis of Chebyshev polynomials which are known to give the best approximation of the exponential to any finite order. Crucially, the Chebyshev polynomials $T_{k}(x)$ can be defined
by the recurrence relation
\begin{equation}
T_{k+1}(x)=2xT_{k}(x)-T_{k-1}(x)\quad,\quad T_{0}(x)=1\quad,\quad T_{1}(x)=x
\label{ChebyshevRecurrence}
\end{equation}
and  form a complete basis for functions on the interval $[-1,1]$. The exponential
time evolution operator can be expanded as
\begin{equation}
e^{-itH}|\Omega\rangle  =e^{-it\bar{H}} J_{0}(at)|\Omega\rangle+e^{-it\bar{H}}2\sum_{k=1}^{\infty}(-i)^{k}J_{k}(at)T_{k}(\Delta H/a)|\Omega\rangle
\end{equation}
where
\begin{equation}
    H=\bar{H}+\Delta H\,, \qquad \bar{H}=\frac{E_{\text{max}}+E_{\text{min}}}{2}\,,
\end{equation}
that is $\bar{H}$ is the average of the maximal and minimal eigenenergies of the truncated Hamiltonian $H$ which is a constant real number,  $J_k$ are the Bessel functions
\begin{equation}
J_{k}(z)=\sum_{l=0}^{\infty}\frac{(-1)^{l}}{l!(k+l)!}\left(\frac{z}{2}\right)^{2l+k}\,,
\label{eq:ExpChebyshev}
\end{equation}
and $a$ is a real number which is larger than the absolute value of all the eigenvalues of the (truncated) Hamiltonian $\Delta H$, but otherwise arbitrary. The use of the constant shift $\bar{H}$ and the multiplicative real number $a$ ensures that the spectrum of the truncated Hamiltonian $\Delta H$ lies in the $[-1,1]$ interval, where the Chebyshev expansion is applicable. The expansion for the exponential can be truncated at an appropriate order to get an approximation for the time evolution operator; it turns out that it is necessary to truncate at a level $n_{max}\gtrsim at_{max}$, with $t_{max}$ the time we aim to reach, which is usually limited by the finite volume and light speed (which is $1$ in our units) to $t\leq L$.

Eq. \eqref{eq:ExpChebyshev} can be rewritten as
\begin{equation}
e^{-itH}|\Omega\rangle  =e^{-it\bar{H}}J_{0}(at)|\Omega\rangle+e^{-it\bar{H}}2\sum_{k=1}^{\infty}(-i)^{k}J_{k}(at)|\Omega\rangle^{(k)}
\label{eq:Chebyshev}
\end{equation}
using
\begin{equation}
|\Omega\rangle^{(k)} =T_{k}(\Delta H/a)|\Omega\rangle
\end{equation}
with $|\Omega\rangle^{(0)}=|\Omega\rangle$. It is immediately obvious that the vectors needed to use (\ref{eq:Chebyshev}) can be computed recursively by rephrasing the recursion relation Eq. \eqref{ChebyshevRecurrence} as
\begin{align*}
|\Omega\rangle^{(1)} & =\frac{1}{a}\left( H|\Omega\rangle^{(0)}-\bar{H}\mathds{1}|\Omega\rangle^{(0)}\right)\\
|\Omega\rangle^{(k+1)} & =2\frac{1}{a}\left( H|\Omega\rangle^{(k)}-\bar{H}\mathds{1}|\Omega\rangle^{(k)}\right)-|\Omega\rangle^{(k-1)}\,,
\end{align*}
which requires only the matrix action of the original Hamiltonian $H$ and a trivial multiplication by the real number $\bar{H}$.

\subsubsection{Solving the Schrödinger equation}

Another way to obtain the time evolution is to solve the time-dependent Schrödinger equation:
\begin{equation}
\label{eq:diffeq_timeevol}
    i\partial_t \ket{\Psi} = H \ket{\Psi}\,.
\end{equation}
This method has the advantage that the Hamiltonian $H$ is allowed to have an explicit time-dependence that does occur in certain physical problems such as the Kibble-Zurek scenario considered in Subsection \ref{subsec:KZ}. Several efficient built-in numerical routines are available to solve a first order linear differential equation in standard mathematical software including MATLAB \cite{MATLAB}, making this method applicable to a wide range of physical problems.

This method has the disadvantage that it is difficult to estimate the duration up to which the solution of the differential equation faithfully describes the actual physical time evolution. Truncation errors limit the validity of this method to a finite time window which can be estimated by simulating the dynamics using different cutoff parameters and observing until which time instant they converge. We note that this limitation is not unique to this specific method of time evolution, and also for a number of physical problems the available window is sufficient to address the relevant questions, as demonstrated in Subsection \ref{subsec:KZ}.

\section{Examples of applications of the CFTCSA to physical problems}
\label{sec:appications}

Here we describe applications of the CFTCSA for relevant perturbations of the critical and tricritical Ising conformal field theories, and the sine-Gordon model. The library of MATLAB \cite{MATLAB} scripts implementing the CFTCSA algorithm described in the previous subsection, as well as CFT data for these models and example calculations can be found online \cite{scripts}. In the case of the first two examples, i.e., the spectral problems of the $E_{8}$ and sine-Gordon QFTs and the form factors of the $E_7$ model, the codes necessary to reproduce all CFTCSA results presented in the paper are fully included as examples. For the other two examples involving time evolution, we refrain from including the full codes due to the complexity of the physical problems, and instead give a simpler but still illustrative code computing time evolution after a global quench in the sine-Gordon model.

\subsection{Spectra of $E_8$ and sine-Gordon model}
\label{ssec:application_spectra}
In the line of its possible applications, let us first present how the CFTCSA calculates the spectrum of two different physical models. The spectral problem is appealing due to its simple formulation and due to the availability of analytical results for the energies. Consequently, the spectral problem benchmarks the accuracy of the numerical method.

\begin{table}[t]
\centering
\begin{tabular}{|c|c|c|c|}
\hline
Volume $m_1 L$ & TBA          & Raw CFTCSA         & Extrapolated CFTCSA \\ \hline\hline
0.075          & -3.490664764718 & -3.490664764706 & -3.490664764727   \\ \hline
0.125          & -2.094420612223 & -2.094420612172 & -2.094420612249   \\ \hline
0.475          & -0.5521585879901 & -0.5521585859634 & -0.5521585879800 \\ \hline
0.6            & -0.4382356381999 & -0.4382356343468 & -0.4382356381806   \\ \hline
0.8            & -0.3314363841477 & -0.3314363756476 & -0.3314363841051   \\ \hline
1.2            & -0.2307543455439 & -0.2307543196153 & -0.2307543454197   \\ \hline
1.6            & -0.1904900446243 & -0.1904899873971 & -0.1904900443767   \\ \hline
2              & -0.1777603739145 & -0.1777602681131 & -0.1777603735367   \\ \hline
4              & -0.2512909490675 & -0.2512902327811 & -0.2512909507354   \\ \hline
7              & -0.4322470994378 & -0.4322437272677 & -0.4322471438901   \\ \hline
9              & -0.5555744641670 & -0.5555676826358 & -0.5555746099325   \\ \hline
12             & -0.7407438075920 & -0.7407286652376 & -0.7407443409462   \\ \hline
\end{tabular}
\caption{Ground state energy $E_0(L)$ for various volume parameters. Data in the second column is calculated from Ref. \cite{1991NuPhB.350..635K}. The third column is obtained by numerical diagonalisation from CFTCSA using 207,809 vectors, while the last column is improved by extrapolating the cut-off dependence.\label{tab:e8e0l}}
\end{table}

The $E_8$ model is formulated by perturbing the Ising conformal field theory with $c=1/2$ by the primary field $\sigma(x)$ of weights $h_\sigma=\bar{h}_\sigma=1/16$ corresponding to magnetisation. This model is integrable and its spectrum is known to consist of $8$ massive particle with masses $m_k$, $k=1,\dots,8$ \cite{Zamolodchikov:1989fp}. The coupling constant $\lambda$ is dimensionful and its exact relation to the mass gap $m_1$ is known \cite{1994PhLB..324...45F}
\begin{equation}
    \lambda_\mathrm{E_8} = \kappa m_1^{15/8}\quad,\quad \kappa = 0.06203236\dots
\label{eq:e8gap}
\end{equation}
so the energy levels can be computed in units of $m_1$, while the volume can be parameterised by the scaling variable $l=m_1L$. The ground state energy in finite volume can be expressed as
\begin{equation}
    E_0(L) =  m_1^2 \mathcal{E}_\mathrm{E_8} L - \frac{\pi\tilde c(l)}{6L}\,,
\label{E0}
\end{equation}
where $\tilde c(l)$ is the so-called vacuum scaling function a.k.a. effective central charge, which behaves as $e^{-l}$ for large volume. The coefficient $\mathcal{E}_\mathrm{E_8}$ is the bulk energy constant which is also exactly known \cite{1994PhLB..324...45F}
\begin{equation}
    \mathcal{E}_\mathrm{E_8} = -\frac{\sin{\pi/30}}{16\sin{\pi/3}\sin{\pi/5}\sin{\pi/15}}\,.
\end{equation}
The $\tilde{c}(l)$ effective central charge was calculated by Klassen and Melzer \cite{1991NuPhB.350..635K} using the thermodynamic Bethe Ansatz (TBA) \cite{Zamolodchikov:1989cf}, which we use to benchmark our numerical results as shown in Table \ref{tab:e8e0l}. We find a remarkable 10-digit accuracy below $l=1$, and a still impressive 5-6 digit agreement up until around $l=10$. 

To illustrate the behaviour of low-lying levels, we present results for the energy level $E_1(L)$ of the first excited state $E_8$ spectrum. This level correspond to the lightest particle with mass $m_1$, and so $E_1(L)-E_0(L)\rightarrow m_1$ as $L\rightarrow\infty$, with finite size corrections which were computed in \cite{1991NuPhB.362..329K} based on the seminal work \cite{Luscher:1985dn} by L{\"u}scher. We compare the predictions for the finite size corrections (in units $m_1=1$) to the CFTCSA results in Table \ref{tab:e8m12l}. 

\begin{table}
\centering
\begin{tabular}{|c|c|c|c|}
\hline
Volume $m_1 L$ & $\Delta m_1$, predicted & $\Delta m_1$, raw CFTCSA & $\Delta m_1$, extrapolated \\ \hline\hline
12.4613        & -2.4860E-03      & -2.4281E-03            & -2.4274E-03                                \\ \hline
13.3811        & -1.0850E-03      & -1.0736E-03            & -1.0729E-03                               \\ \hline
14.2486        & -5.0140E-04      & -4.9909E-04            & -4.9846E-04                                  \\ \hline
15.0722        & -2.4220E-04      & -2.4194E-04            & -2.4139E-04                                  \\ \hline
15.8582        & -1.2130E-04      & -1.2154E-04            & -1.2111E-04                                  \\ \hline
16.6114        & -6.2680E-05      & -6.2983E-05            & -6.2714E-05                                  \\ \hline
\end{tabular}

\caption{Finite-size corrections to the lowest-lying particle's energy in the $E_8$ spectrum. KM stands for the analytical results of Klassen and Melzer \cite{1991NuPhB.362..329K}. \label{tab:e8m12l}}
\end{table}

Another example we consider is the sine-Gordon model with the Hamiltonian
\begin{equation}
    H_{\text{sG}}=\int \text{d}x \left[\frac{1}{2}\left(\partial_t\varphi(x,t)\right)^2
    +\frac{1}{2}\left(\partial_x\varphi(x,t)\right)^2
    -\lambda_{\text{sG}} :\cos\beta\varphi(x,t):\right]
\label{eq:homsGHam}
\end{equation}
which is a relevant perturbation of the massless boson CFT with $c=1$ for $\beta^2<8\pi$ (c.f. Appendix \ref{sec:boson} for more detailed definitions). Due to the periodic potential, the boson can be considered as an angular variable, with the radius of the compactification circle in CFT conventions given by 
\begin{equation}
    R=\frac{\sqrt{4\pi}}{\beta}\,.
\end{equation}
We work in the attractive regime $\beta<\sqrt{4\pi}$ where the force between solitons and antisolitons is attractive, leading to the existence of breather bound states, and use the mass $m_1$ of the lightest breather to define our units. Due to integrability, the exact relation of the coupling $\lambda$ to the mass scale $m_1$ is also known \cite{1995IJMPA..10.1125Z}, which allows us to rescale all quantities to units $m_1=1$ via
\begin{equation}
\lambda_{\text{sG}}=m_1^{2-2\Delta}\frac{\Gamma\left(\Delta\right)}{\pi\Gamma\left(1-\Delta\right)}\left[\frac{\sqrt{\pi}\Gamma\left(\frac{1}{2-2\Delta}\right)}{4\sin\left(\frac{\pi\Delta}{2-2\Delta}\right)\Gamma\left(\frac{\Delta}{2-2\Delta}\right)}\right]^{2-2\Delta}\,,\qquad \Delta=h=\bar{h}=\frac{1}{2R^2}\,.\label{eq:lambdaQM}
\end{equation}
In addition, ``numerically exact'' predictions for the finite volume ground state energy can be computed from the NLIE method \cite{1991JPhA...24.3111K,1992PhRvL..69.2313D}, using tools developed in \cite{1998PhLB..430..264F}, which are compared to the CFTCSA results in Table \ref{tab:sGe0l}. 

\begin{table}
\centering
\begin{tabular}{|c|c|c|c|}
\hline
Volume $m_1 L$ &  Num. exact (NLIE)          & Raw CFTCSA    & Extrapolated CFTCSA     \\ \hline\hline
0.5  & -1.063251174 & -1.0632500013 & -1.0632501814\\ \hline
1.0  & -0.6107173129 & -0.6107108772 & -0.6107129149\\ \hline
1.5  & -0.5637456919 & -0.5637275871 & -0.56373601\\ \hline
2.0  & -0.6343390247 & -0.6343022845 & -0.6343253386\\ \hline
2.5  & -0.7480121071 & -0.7479459813 & -0.7479963236\\  \hline
3.0  & -0.8784116537 & -0.8783099401 & -0.8784052346\\ \hline
3.5  & -1.0158445810 & -1.0156954441 & -1.0158588928\\ \hline
4.0  & -1.1564966832 & -1.1562870673 & -1.1565478948\\ \hline
4.5  & -1.2987298702 & -1.2984468181 & -1.2988407192\\ \hline
5.0  & -1.4417811458 & -1.4414115972 & -1.4419811557\\ \hline
\end{tabular}
\caption{The ground state energy $E_0(l)$ for various volume parameters for the sine-Gordon model. Data in the second column is from Ref. \cite{1991NuPhB.350..635K}. The third column is obtained by numerical diagonalisation from CFTCSA with $E_{c}=40$ and the compactification radius $R=2$ resulting in
a truncated Hilbert space of dimension $5,320,750$, while the last column is improved by extrapolating the cut-off dependence.\label{tab:sGe0l}}
\end{table}

Extending to larger volumes we can test the bulk contribution to the ground state energy, and the masses $m_1$ and $m_2$ of the first two breathers, as shown in Table \ref{tab:sGlargevol}. The asymptotic behaviour of the ground state energy is given by \cite{Destri:1990ps}
\begin{equation}
    E_0(L)=m_1^2\mathcal{E}_{\text{sG}} L+O\left(e^{-m_1 L}\right)\quad,\quad \mathcal{E}_{\text{sG}}=-\left(8\sin\frac{\pi}{2R^2-1}\right)^{-1}\,,
\end{equation}
while in our units $m_1=1$ and the exact mass of the second breather is given by \cite{Zamolodchikov:1978xm}
\begin{equation}
    m_2=2m_1\cos\frac{\pi}{2(2R^2-1)}\,,
\end{equation}
when $R^2>3/2$, that is, when the second breather is present in the spectrum.
We also computed the leading order finite size corrections ($\mu$-terms) for the second breather masses following \cite{1991NuPhB.362..329K}, which notably improves the agreement for the second breather\footnote{For the first breather, the deviations between the asymptotic mass and the TCSA result are dominated by the truncation errors, indicated by the fact that it increases with volume.}. 

\begin{table}[h]
\centering%
\begin{tabular}{|c|c|c||c|c||c|c|c|}
\hline 
 &  $E_{0}$ & As. ex. $E_{0}$  & $m_{1}$ & As. ex. $m_{1}$ & $m_{2}$ & As. ex. $m_{2}$+$\mu$ terms & As. ex. $m_{2}$\tabularnewline
\hline 
\hline 
$Lm_{1}=12$ &  -3.4536 & -3.4572 & 1.0001 & 1 & 1.9376 & 1.9350 & 1.9499\tabularnewline
\hline 
$Lm_{1}=18$ &  -5.1751 & -5.1857 & 1.0005 & 1 & 1.9471 & 1.9460 & 1.9499\tabularnewline
\hline 
\end{tabular}
\caption{\label{tab:sGlargevol}
The CFTCSA energies for the ground state $E_0$
and the first two excited state $m_{1}$ and $m_{2}$ for the SG model when $Lm_{1}=12$
and $Lm_{1}=18$. The numerical values are compared to the asymptotically exact (As. ex.) results and the breather masses to  analytic expression incorporating the (strongest volume dependent) Lüscher corrections as well. The cut-off parameter $E_{c}=40$ and the compactification radius $R=2$ resulting in a truncated
Hilbert space of dimension 5,320,750.}
\end{table}

\subsection{Form factors in the $E_7$ model}

The $E_7$ model is given by tricritical Ising model with central charge $c=7/10$ perturbed by the operator of conformal dimension $h=\bar{h}=1/10$. It is an integrable field theory \cite{MC,FZ}, with the mass gap related to the coupling $\lambda$ as \cite{1994PhLB..324...45F}
\begin{equation}
  m_1=|g|^{5/9} 3.7453728362\dots \,.
  \label{e7:masscouplingrelation}
\end{equation}
The positive/negative signs of the coupling correspond to paramagnetic/ferromagnetic phases. The CFTCSA was applied to this model in the recent work \cite{2021arXiv210909767C} in order to support and verify the construction of form factors of order/disorder operators, which were subsequently used to compute dynamical structure factors. The example scripts included in \cite{scripts} demonstrate the evaluation of the one-particle form factors of the leading and subleading magnetisation operators in the paramagnetic regime $g>0$. The results are summarised and compared to predictions of the form factor bootstrap in Table \ref{tab:1ptTCSAsigma}; note that the results are shown with the precision attained in \cite{2021arXiv210909767C} which used a cut-off at chiral level $20$ resulting in a truncated Hilbert space with $623,552$ states. For further details the interested reader is referred to \cite{2021arXiv210909767C}.

\begin{table}[ht]
	\centering
	\bgroup
    \setlength{\tabcolsep}{0.5em}
	\begin{tabular}{|c|c|c|c|c|}
		\hline
		Particle & $|F_i^{\sigma}|$ & $|F_i^{\sigma}|$ &$|F_i^{\sigma'}|$ &$|F_i^{\sigma'}|$ \\
		\hline
		 & bootstrap & CFTCSA & bootstrap & CFTCSA\\
		\hline
		$1$ & $0.71043$ & $0.71017$ & $2.05592$ & $2.04971$\\
		\hline
		$3$ & $0.25232$ & $0.25165$ & $1.71395$ & $1.68341$ \\
		\hline
	\end{tabular}
	\egroup
	\caption{One-particle form factors of the magnetisation operators: predictions of the form factor bootstrap compared to CFTCSA.\label{tab:1ptTCSAsigma} }
\end{table}

\subsection{Inhomogeneous quenches in the sine-Gordon field theory}
\label{InhomSGProblem}

Recently, the chirally factorised TCSA was applied to inhomogeneous quantum quenches in the sine-Gordon theory in \cite{2021arXiv210906869H}. In this quench protocol we considered the time evolution by the homogeneous sine-Gordon Hamiltonian \eqref{eq:homsGHam}, starting from an initial state defined as the ground state of an inhomogeneous sine-Gordon Hamiltonian:
\begin{equation}
H_{\text{inhom}}=H_{\text{sG}}-\int\text{d}x\,\partial_{x}\varphi(x,t)\,j'(x)\,,
\label{eq:InHomQMHamGeneral}
\end{equation}
where $\partial_{x}\varphi$ is the spatial derivative of the fundamental bosonic field in the theory and $j(x)$ is a static source term or external field and $j'$ denotes its spatial derivative. 
The main computational challenge is attributed to the above Hamiltonian and to determining its ground state. Due to the inhomogeneous term in \eqref{eq:InHomQMHamGeneral}, the full Hamiltonian is no longer translation invariant, in other words, the Hamiltonian connects different momentum sectors of the Hilbert space, whose separate treatments are hence not possible. As a consequence, very large Hilbert spaces had to be handled whose largest dimensions were of the order of $10^6$. 

The determination of the ground state of the inhomogeneous Hamiltonian Eq. \eqref{eq:InHomQMHamGeneral} can be carried out using the CFTCSA method with some additional considerations. Here we review these aspects briefly and refer the interested reader to Appendices  \ref{app:vertexopmatrix} and \ref{AppB4} for details. The most essential consideration is to classify the terms in the Hamiltonian \eqref{eq:InHomQMHamGeneral} according to their action in terms of momentum sectors. The integrated vertex operators (i.e. the cosine interaction) are translationally invariant and therefore block diagonal in terms of momentum sectors, whereas the derivative field  $\partial_{x}\varphi$ integrated with the inhomogeneous classical field $j'(x)$ connects sectors with different momentum. This must be carefully encoded in the corresponding operator descriptors, for which it is useful to organise the Hilbert Space Descriptors according to momentum sectors, or alternatively use a collection (array) of separate descriptors for each momentum sector (c.f.  \ref{AppB4}). Then the application of the inhomogeneous Hamiltonian on a vector can easily be specified in the CFTCSA language and the determination of the ground state boils down to the same spectral problem we already discussed.

Evaluation of the subsequent time evolution is much less demanding in computing power since it can be performed separately in each momentum sector due to the translation invariance of the post-quench Hamiltonian \eqref{eq:homsGHam}. In particular, decomposing the inhomogeneous initial state $|0\rangle_j$ (ground state of \eqref{eq:InHomQMHamGeneral}) as
\begin{equation}
    |0\rangle_j= \bigoplus_{k} |0\rangle^{(k)}_j
\end{equation}
according to various momentum sectors labelled by the index $k$, the time evolution with the homogeneous Hamiltonian $H_\text{sG}$ can be written as
\begin{equation}
\label{sGTimeEvolSectors}
    e^{-itH_\text{sG}}|0\rangle_j= \bigoplus_{k} e^{-itH_\text{sG}}|0\rangle^{(k)}_j=\bigoplus_{k} e^{-itH^{(k)}_\text{sG}}|0\rangle^{(k)}_j\,,
\end{equation}
where $H^{(k)}_\text{sG}$ is the homogeneous sine-Gordon Hamiltonian restricted onto a single momentum sector. The RHS of Eq. \eqref{sGTimeEvolSectors} can then be evaluated by a straightforward application of the Chebyshev-Bessel method reviewed in Subsection \ref{subsec:CBmethod}. The expectation values of local operators such as $\partial_x \varphi$ can then be performed by a straightforward application of Eq. \eqref{eq:matelem_general}.

In the following, demonstrate some results of Ref. \cite{2021arXiv210906869H} focusing on the properties of the inhomogeneous ground state and also provide a benchmark for the validity of our results.
The relevant physics behind the initial state can be understood as follows. The topological charge density $\rho$ and the charge itself $Q$ (corresponding to the number of solitons minus the anti-solitons) are given as
\begin{equation}
\rho(x,t)=\frac{\beta}{2\pi}\partial_x\varphi(x,t)\,, \qquad Q=\int \text{d}x\rho(x,t)\,.
\label{rhodxphi}
\end{equation}
Note that in \eqref{eq:InHomQMHamGeneral}, the topological charge density is coupled to the inhomogeneous external source $j'(x)$, which can, therefore, also be regarded as an external chemical potential for solitons. We denote the ground state of the above Hamiltonian by $|0\rangle_{j}$ and the ground state expectation values as $\langle...\rangle_j$. In particular, we define the expectation value of the field $\varphi$ as
\begin{equation}
\langle \varphi(x,t)\rangle_j:=\int_{-L/2}^{x}\text{d}x' \langle\partial_{x'}\varphi(x',t)\rangle_j\,,
\label{DefForVarphi}
\end{equation}
since unlike $\partial_x\varphi$, the field $\varphi$ is not a well-defined quantity in the sine-Gordon QFT.
In Ref. \cite{2021arXiv210906869H}, $j'$ was chosen to be a localised inhomogeneity expressed by a Gaussian bump
\begin{equation}
j'(x)=\frac{A }{\sqrt{2\pi}\sigma m_1}\exp\left(-\frac{x^{2}}{2\sigma^{2}}\right)-\frac{A}{\ell}\,\text{erf}\left(\frac{\ell}{2\sqrt{2}\sigma m_1}\right)\,,\quad \int_{-L/2}^{L/2}\text{d}x\,j'(x)=0,
\end{equation}
centred at the middle
of the interval $[-L/2,L/2]$. Here $\ell$ denotes the dimensionless volume (length) of the system $\ell=m_1L,$ and $m_1^{-1} A$ and $m_1 \sigma$ are dimensionless parameters controlling the amplitude
and the width of the Gaussian bump. This external field imposes a neutrality condition which is achieved by subtracting the constant term $A/\ell\,\text{erf}\left(\ell/(2\sqrt{2} m_1 \sigma)\right)$ from the Gaussian bump and which guarantees that the ground state of \eqref{eq:InHomQMHamGeneral} is in the topologically neutral sector.

\begin{figure}
\begin{subfigure}[b]{0.48\textwidth}
\centering
\includegraphics[width=\textwidth]{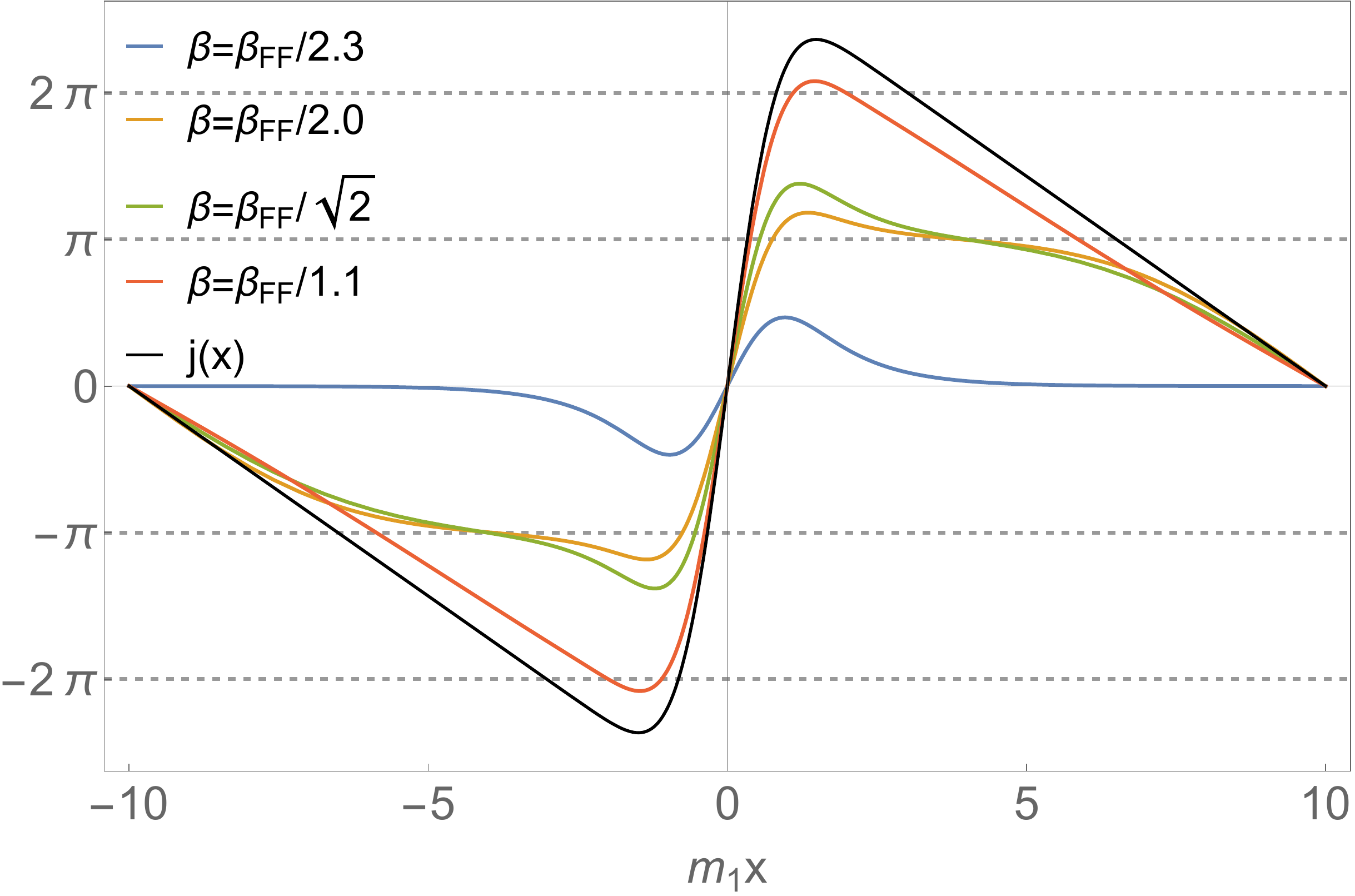}
\caption{\centering
$\beta\langle\varphi(x)\rangle_{j}$}
\end{subfigure}\hfill
\begin{subfigure}[b]{0.48\textwidth}
\centering
\includegraphics[width=\textwidth]{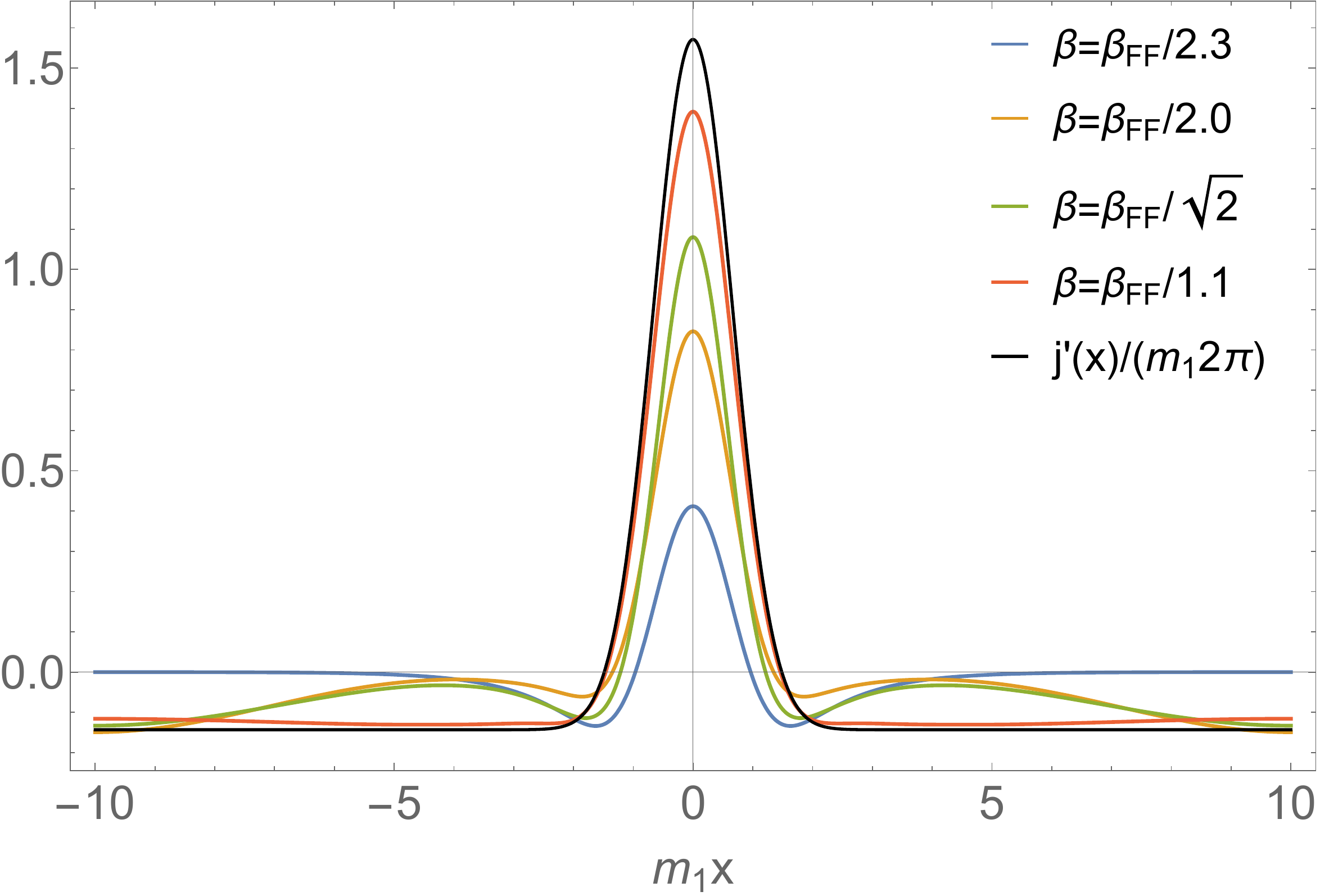}
\caption{\centering
$m_{1}^{-1} \langle \rho(x)\rangle_{j}$}
\end{subfigure}
\caption{\label{figW450GSTCSA}
Field profiles for different values of the coupling $\beta$: 
(a) the QFT expectation values $\langle\beta\varphi(x)\rangle_{j}$ and (b) the corresponding topological charge densities $m_{1}^{-1}\langle\rho(x)\rangle_{j}$ in the quantum case. The parameters are $\ell=m_1L=20$, $m=m_1$, $\beta_\text{FF}A/m_1=18$, $m_1 \sigma=2/3$; different $\beta$ values are shown with different colour and $\beta_\text{FF}=\sqrt{4\pi}$.
The CFTCSA profiles $\langle\rho(x)\rangle_{j}$ were extrapolated using  cut-offs $E_{c}=24,26,28$ and $30$, and the corresponding profiles $\beta\langle\varphi(x)\rangle_{j}$ were obtained by spatial integration of $2\pi\langle\rho(x)\rangle_{j}=\beta\langle\partial_{x}\varphi(x)\rangle_{j}$, fixing the zero mode by requiring the result to vanish at the origin $x=0$.
}
\end{figure}

The initial state of the inhomogeneous Hamiltonian displays an interesting transition when changing either the amplitude $A$ of the external field or the interaction parameter $\beta$ of the sine--Gordon model. The transition is reflected by changes in the initial profile of the field $\varphi$ and the topological charge density ($\rho\propto \partial_{x}\varphi$), as well as jumps in the field zero mode \cite{2021arXiv210906869H}. In particular, Figure \ref{figW450GSTCSA} demonstrates that fixing the amplitude $A$ of the external field and varying the interaction parameter $\beta$ steep transitions happen in the sine-Gordon theory which can be associated with the suddenly enhanced soliton content of the initial state $|0\rangle_{j}$. 

\begin{figure}
\begin{subfigure}[b]{0.48\textwidth}
\centering
\includegraphics[width=\textwidth]{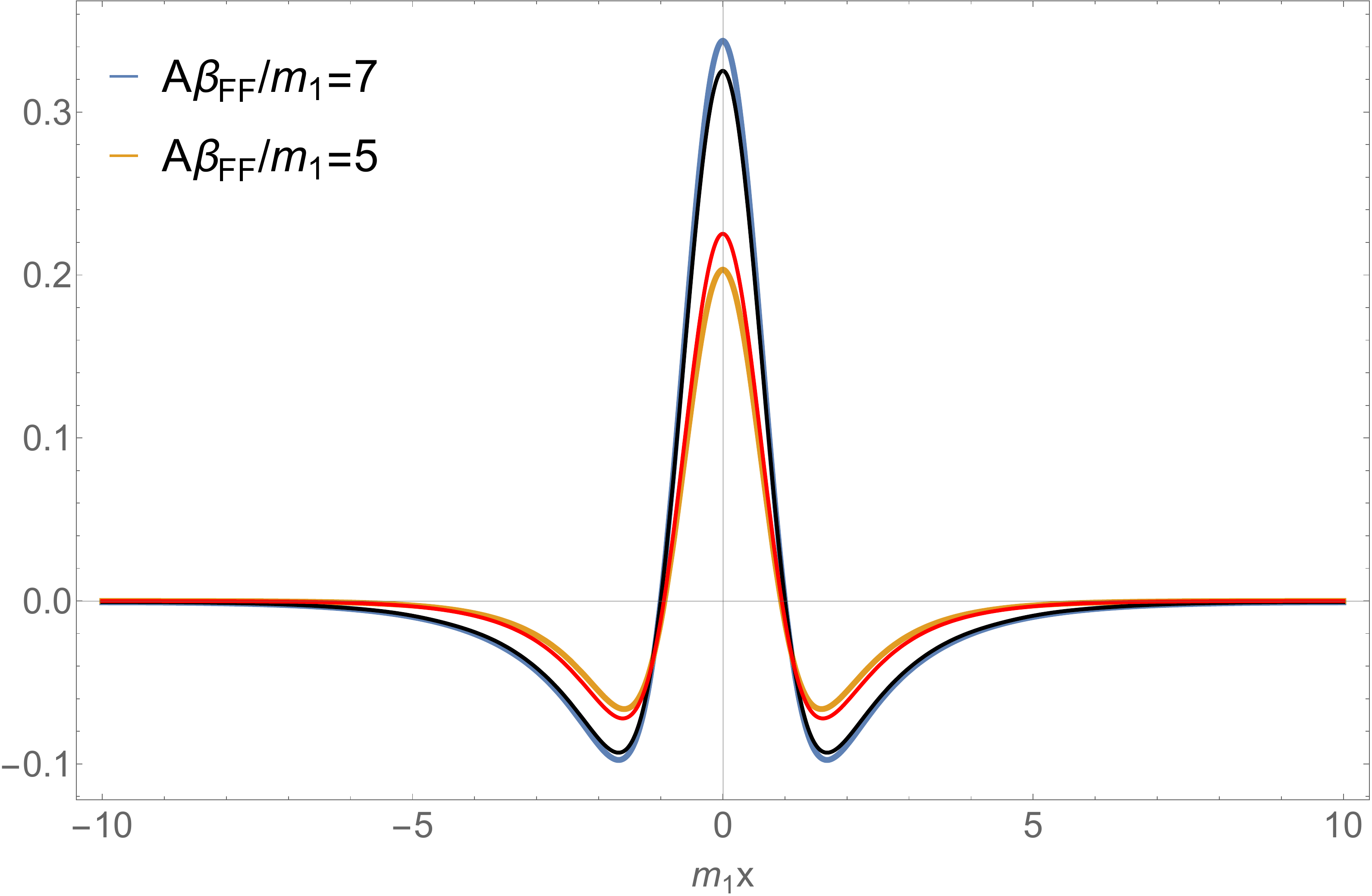}
\caption{\centering
$m_{1}^{-1} \langle\rho(x)\rangle_{j}$ (small amplitudes)}
\end{subfigure}\hfill
\begin{subfigure}[b]{0.48\textwidth}
\centering
\includegraphics[width=\textwidth]{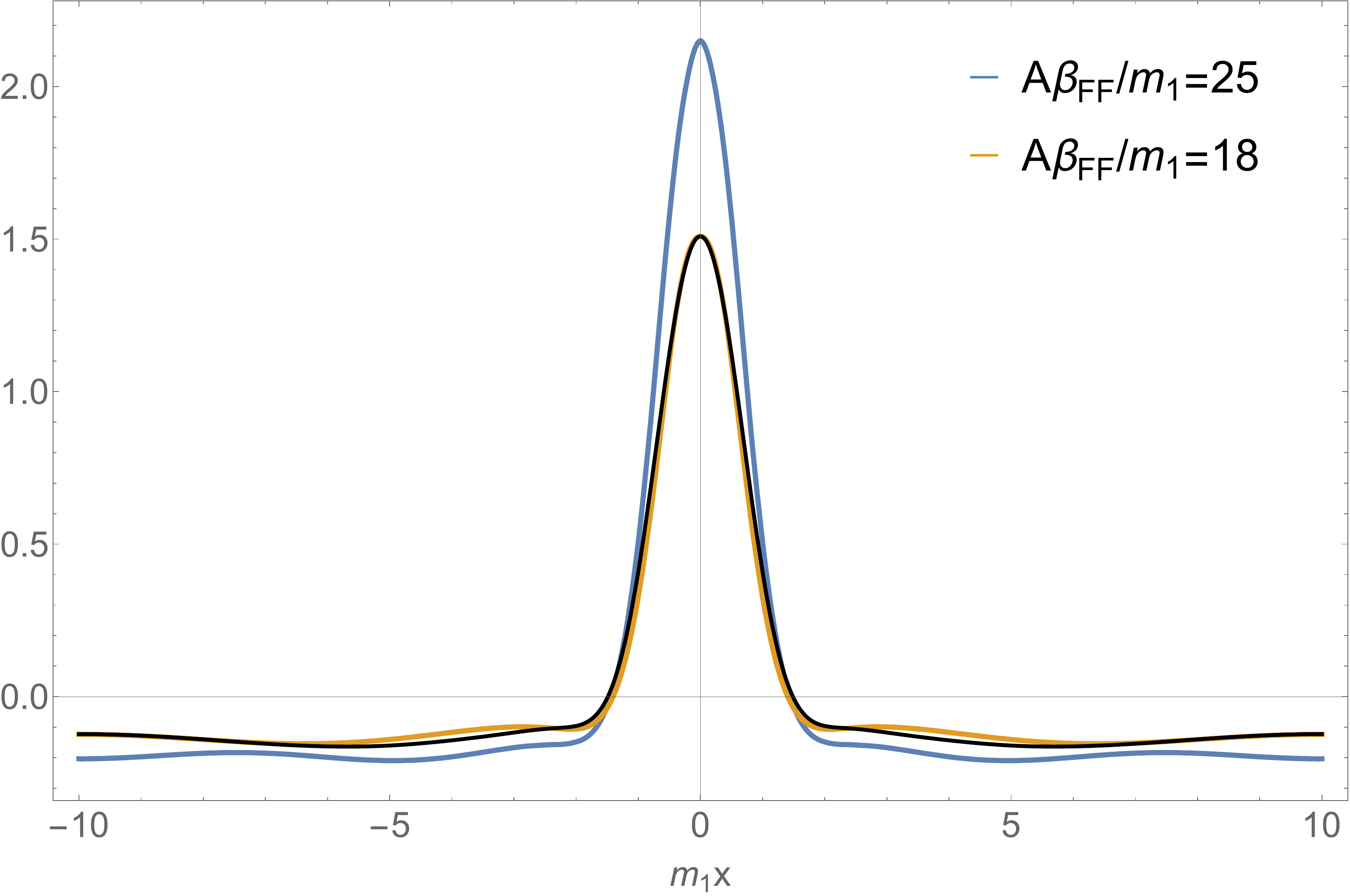}
\caption{\centering
$m_{1}^{-1} \langle\rho(x)\rangle_{j}$ (large amplitudes)}
\end{subfigure}
\caption{The QFT expectation value of $\langle\rho(x)\rangle_{j}$ at the free fermion point ($\beta=\beta_{\text{FF}}$) of the sine-Gordon model for four different amplitudes of the external field: $A \beta_\text{FF}/m_1=5$, $A \beta_\text{FF}/m_1=7$, and $A \beta_\text{FF}/m_1=18$, $A \beta_\text{FF}/m_1=25$. The bump-width in the external field is $m_1\sigma=2/3$ in all cases. The coloured continuous lines correspond to  results from exact free fermion computation,  and for 
$A \beta_\text{FF}/m_1=7$, and $A \beta_\text{FF}/m_1=18$ the CFTCSA profiles with continuous black lines are also shown. For the CFTCSA curves extrapolation was used based on the data with cut-offs $E_{c}=24,26,28$
and $30$. The mass scale $m_1$ is understood as twice the fermion mass $M$. In Subfigure (a) the rescaled Klein--Gordon ($\beta=0$) initial profile $\langle\partial_x \varphi(x)\rangle_j/\sqrt{\pi}$ for $A \beta_\text{FF}/m_1=5$ is also displayed with continuous red line.}
\label{figTCSAFF}
\end{figure}

The same qualitative transition can, nevertheless, be observed when instead of fixing $A$ and changing the interaction $\beta$, the amplitude $A$ is varied and the interaction $\beta$ is kept fixed. Here we demonstrate this fact by focusing on the (topological) charge density of the sine-Gordon theory  solely at the free fermion point (for discussion including other $\beta$ values c.f. \cite{2021arXiv210906869H}). As is well known, the theory is equivalent to the free massive Dirac field theory when $\beta=\beta_{\text{FF}}=\sqrt{4\pi}$, allowing alternative treatments to the problem which can be used  to crosscheck the validity of our CFTCSA results. The transition is displayed in Figure \ref{figTCSAFF}, which compares the results of exact free-fermion computations and the truncation extrapolated curves obtained from the CFTCSA. The excellent match between the curves is a solid justification for the accuracy of the CFTCSA, especially since sine-Gordon TCSA is known to be much less convergent when $\beta$ approaches the free fermion value $\beta_\mathrm{FF}$, and even has ultraviolet divergences in the repulsive regime $\beta\geq\beta_\mathrm{FF}$, which however do not preclude its use \cite{1998PhLB..444..442F,1999NuPhB.540..543F}. At the free fermion point only the vacuum energy is divergent, which cancels from the quantities shown in Figure \ref{figTCSAFF}.  

The transition observed in the initial state can be understood via semi-classical considerations discussed in Ref. \cite{2021arXiv210906869H}. Additionally, below the first transition point (i.e. for small $\beta$ or $A$), the profiles were found to be very similar to profiles obtained from a Klein--Gordon theory with the same scalar particle mass. However, beyond the first transition point, quantum profiles  develop features of the analogous free Dirac fermion problem for high enough amplitudes of the initial external field, at least in the investigated parameter regime. It is important to stress that the above transition is present also in the Dirac theory when changing the amplitude of the external field as shown by Figure \ref{figTCSAFF}. The emergence or absence of the fermionic features of the inhomogeneous initial states also profoundly change the subsequent time evolution of the profiles with the homogeneous sine-Gordon Hamiltonian, which can be evaluated using the Chebyshev-Bessel method described in Subsection \ref{subsec:CBmethod}, and is discussed in detail in Ref. \cite{2021arXiv210906869H}.

\subsection{Kibble-Zurek scaling in Ising field theory}\label{subsec:KZ}

Another recent application of our improved numerical method was the investigation of the Kibble--Zurek scaling in the Ising field theory \cite{2020ScPP....9...55H}. Here we discuss the problem from the point of view of the algorithmic realisation to present the performance of the CFTCSA in modelling complex non-equilibrium dynamics.

The Kibble--Zurek (KZ) scaling \cite{1976JPhA....9.1387K,1980PhR....67..183K,1985Natur.317..505Z,1996PhR...276..177Z} captures the universal features of non-equilibrium dynamics when a system is driven slowly across a continuous phase transition. In a system near criticality time and length scales of the dynamics diverge,  giving rise to the phenomenon called critical slowing down. As a result, it is impossible to drive a system through a critical point in a completely adiabatic way: near the critical point any slow change is fast compared to the infinitely slow reaction time of the system. Therefore, the fully ordered state cannot be obtained by slowly cooling the sample below its critical temperature. The deviation from complete order is signalled by the presence of defects, and for a sufficiently slow process, the density of defects depends universally on the cooling rate, which is captured by the Kibble--Zurek scaling. A completely analogous phenomenon occurs in quantum critical systems, where the critical point is crossed in the ground state of the model by varying some coupling parameter instead of the temperature \cite{2005PhRvL..95j5701Z,2005PhRvL..95x5701D,2005PhRvB..72p1201P}.

\begin{figure}[t]
\centering\includegraphics[width=0.6\textwidth]{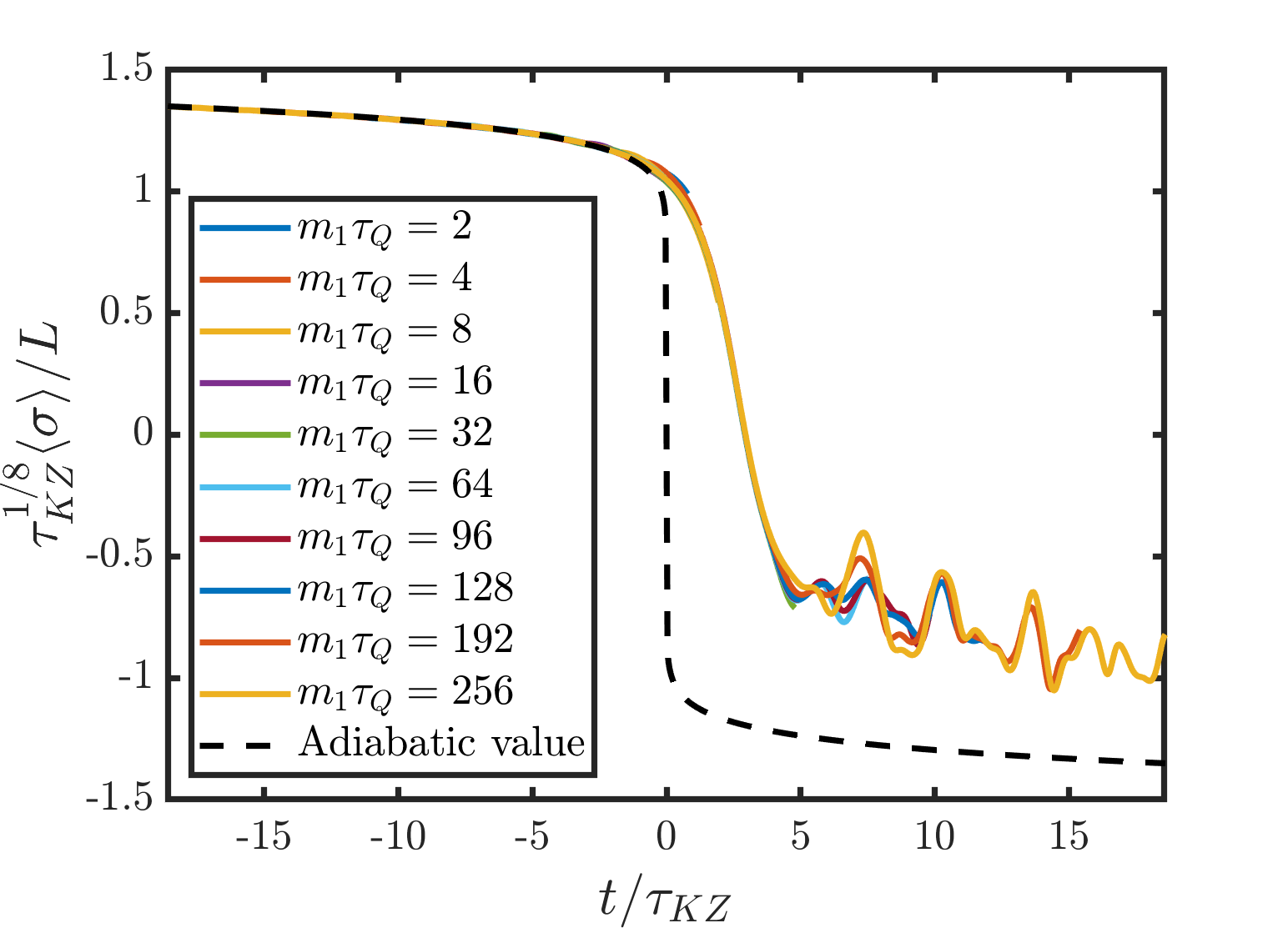}
\caption{CFTCSA modelling the Kibble--Zurek mechanism using the volume parameter $m_1L = 50$. Dashed lines denote the adiabatic value $\expval{\sigma}_\mathrm{ad}$, while continuous lines illustrate extrapolated CFTCSA results for different ramp times $\tau_\mathrm{Q}$. The different curves scale on top of each other outlining the scaling function $F_\sigma(t/\tau_\mathrm{KZ})$.\label{fig:kzm_ex}}
\end{figure}

The KZ scaling applies to more general quantities beyond the density of defects. Let us consider a perturbation of a quantum critical point (QCP) by some operator with scaling dimension $\Delta.$ The strength of the perturbation is characterised by a coupling constant $\delta$ with $\delta=0$ corresponding to the critical point. Assume that we prepare the system in its ground state and drive it through its QCP by changing $\delta$ in time, i.e. by performing a ramp: 
	\begin{equation}
	\label{eq:ramp}
	\delta = \delta_0\frac{t}{\tau_\mathrm{Q}}\,,
	\end{equation}
where $\tau_\mathrm{Q}$ is the rate of the quench. Then we can identify the time and length scales corresponding to departure from adiabatic behaviour:
	\begin{equation}
	\label{eq:tKZ}
	\tau_\mathrm{KZ}\equiv (\nu z)^{\frac1{\nu z+1}} \left(\frac{\tau_\mathrm{Q}}{\delta_0}\right)^{\frac{\nu z}{\nu z+1}}\,,\qquad \xi_\mathrm{KZ}\equiv\xi(-\tau_\mathrm{KZ}) \propto \left(\frac{\tau_\mathrm{Q}}{\delta_0}\right)^{\frac{\nu z}{\nu z+1}}\propto\tau_\mathrm{KZ}\,,
	\end{equation}
where $\nu$ and $z$ are the critical exponents: $\nu$ relates the divergence of the correlation length to the distance from the critical point $\delta$, while $z$ characterises the massless dispersion relation $E\propto k^z$ (in a relativistic theory $z=1$). In the KZ time window
\begin{equation}
-\tau_\mathrm{KZ}<t<\tau_\mathrm{KZ}
\label{eq:KZwindow}
\end{equation}
the various physical quantities are subject to the KZ scaling. This is expressed by rescaling them with $\tau_\mathrm{KZ}$ and $\xi_\mathrm{KZ}$. For instance, the dynamical expectation value of a local operator in a translationally invariant model takes the form
\begin{equation}
	\label{eq:KZ_fintimsc}
	\expval{\mathcal{O} (x,t)} = \xi_\mathrm{KZ}^{-{\Delta_\mathcal{O}}}F_\mathcal{O}(t/\tau_\mathrm{KZ})\,,
\end{equation}
where $\Delta_\mathcal{O}=h_\mathcal{O}+\bar{h}_\mathcal{O}$ is the scaling dimension of the operator $\mathcal{O}$ and $F_\mathcal{O}$ is a universal scaling function, and we assumed translational invariance.

Using the notation introduced in Section~\ref{sec:implementation} we can express the physical problem of the Kibble--Zurek mechanism using the CFTCSA. The conformal Hamiltonian describes a quantum critical system, and the slow ramp is realised by adding a perturbation with a time-dependent coupling constant. On the algorithmic level, the time evolution is obtained by numerically solving Eq. \eqref{eq:diffeq_timeevol}. The CFTCSA package provides an efficient evaluation of the action of the time-dependent Hamiltonian on a general vector using Eqs. \eqref{eq:localOpAction} and \eqref{eq:confHamaction}. This action can then be fed as input into an appropriate differential equation solver, and in \cite{2020ScPP....9...55H} we used the routine `ode45' (a fifth-order Runge--Kutta method based on Ref. \cite{DORMAND198019}) which is part of MATLAB. From the time-dependent state vector the expectation value Eq. \eqref{eq:KZ_fintimsc} can then be computed using \eqref{eq:matelem_general}.

Taking the $E_8$ direction of the Ising Field Theory described in  Subsection~\ref{ssec:application_spectra} as an example, the perturbing operator $\mathcal{V}$ corresponds to the magnetisation operator $\sigma$ with $h_\sigma = \bar{h}_\sigma = 1/16$. In the ground state the dependence of its expectation value on the coupling constant $h$ is exactly known \cite{1998NuPhB.516..652F}, and yields the adiabatic value $\expval{\sigma}_\mathrm{ad}$. The rescaled dynamical one-point functions $\expval{\sigma(t/\tau_\mathrm{KZ})}$ can be computed using the procedure described in the previous paragraph. Figure \ref{fig:kzm_ex} illustrates that the numerical method is able to capture the nontrivial dynamical behaviour behind the Kibble--Zurek scaling. Moreover, the numerically obtained curves collapse on top of each other even beyond $t/\tau_\mathrm{KZ}=1$, indicating that the KZ scaling is valid in a broader time window than the conservative estimate \eqref{eq:KZwindow}.

\section{Summary}
In this work we presented an implementation of the Truncated Conformal Space Approach for quantum field theories specified as relevant perturbations of $1+1$ conformal field theories, with periodic boundary conditions. Our implementation exploits the chiral factorisation property of conformal field theory to optimise computing performance in terms of the available dimension of the Hilbert space, which is important to attain sufficient numerical precision for many applications. We demonstrated the power of the method using examples from recent works of research, and also made accessible the MATLAB scripts implementing the method together with some explicit examples \cite{scripts}.   

\section*{Acknowledgements}
The authors are grateful to R. Horv{\'a}th and M. L{\'a}jer for their useful pieces of advice which helped us implement the improved numerical method. We are also grateful to G. Feh{\'e}r for his early stage contribution, and to M. Lencs{\'e}s for applying the algorithm to the tricritical Ising model. G. T. was also partially supported by the  National Research, Development and Innovation Office (NKFIH) through the OTKA Grant K 138606, and also within the Quantum Information National Laboratory of Hungary. D. X. H. acknowledges support from ERC under Consolidator grant number 771536 (NEMO). 

\appendix
\section{CFT data for minimal models}\label{sec:Virasoro}

For the case of minimal models of conformal field theory with central charge \begin{equation}
    c=1-\frac{6}{p(p+1)}\quad,\quad p=3,4,5,\dots
\end{equation}
the chiral algebra is the Virasoro algebra 
\begin{equation}
\left[L_n,L_m \right]=(n-m)L_{n+m}+\frac{c}{12}n(n^2-1)\delta_{n,-m}   
\label{eq:Vir_commutator}
\end{equation}
and the irreducible representations $V_\mathcal{R}$ are just labelled by the corresponding conformal weight $h_\mathcal{R}$, and so they are simply labelled as $V_h$ from now on. These modules can be spanned by vectors of the form\footnote{Note that not all such vectors are independent due to the existence of singular vectors, to which we return later.}
\begin{align}
    &L_{-n_1}\cdots L_{-n_k}\ket{h}\quad ,\quad n_1\geq\dots,\geq n_k\in\mathbb{Z}_+\nonumber\\
    &L_m\ket{h}=0 \quad m\in\mathbb{Z}_+\quad ,\quad L_0\ket{h}=h\ket{h}\,,
\label{eq:Virasoro_descendant}
\end{align}
which have descendant levels $N=n_1+\dots+n_k$. 

The operator algebra structure constants were computed in \cite{1985NuPhB.251..691D,1985PhLB..154..291D} for CFTs with diagonal ($A$ type) partition functions, while for those with $D$ type partition functions they can be found in the papers 
\cite{1989PhLB..225..357P,1995NuPhB.438..347P} and \cite{2000NuPhB.579..561R}. We note that the structure constants are often given in terms of fields which are not normalised according to CFT conventions, and so they must be redefined accordingly to be consistent with exact mass gap relations such as \eqref{eq:e8gap} or \eqref{e7:masscouplingrelation} used to set the units.

\subsection{Three-point couplings of descendant fields}
The chiral three-point matrices  $\mathcal{B}^\mathcal{O}(\mathcal{R}(\Phi_1),N_1,\mathcal{R}(\Phi_2),N_2)$ can be computed from the knowledge of a trilinear function $\mathbb{T}:V_{h_1}\otimes V_{h_3}\otimes V_{h_2} \rightarrow \mathbb{C}$, which describes how to reduce matrix elements of descendant fields to the primary ones. For minimal models, specifying each representation by its weight leads to the simplified notation of the chiral three-point matrices $\mathcal{B}^\mathcal{O}(h_1,N_1,h_2,N_2)$, which is the notation used in the rest of this appendix.

The function $\mathbb{T}$ can be constructed recursively using conformal Ward identities with a method first used in the context of TCSA calculations in \cite{1997NuPhB.489..557K}, and are available in print together with their derivation in \cite{2006NuPhB.744..358K}\footnote{The results of \cite{2006NuPhB.744..358K} contain the extension of the procedure to the case of $N=1$ superconformal symmetry. Here we only make use of the Virasoro part, but the algorithm described in this section can be extended to the superconformal case, as was done in \cite{2006NuPhB.744..358K}}. Here we briefly describe the construction, omitting the derivation which can be found in \cite{2006NuPhB.744..358K}. Every argument of $\mathbb{T}$ can be written in the form \eqref{eq:Virasoro_descendant}, and the matrix elements can be reduced as follows:
\begin{enumerate}
    \item First the left vector is reduced to a primary one by successive application of the relation
    \begin{equation}
        \mathbb{T}\left( L_{-n}\ket{A},\ket{B},\ket{C}\right)=\mathbb{T}\left( \ket{A},\ket{B},L_{n}\ket{C}\right)
        +\sum_{k=-1}^n \left({n+1}\atop{k+1}\right)\mathbb{T}\left( \ket{A},L_{k}\ket{B},\ket{C}\right)\,,
    \label{eq:Vir_left}\end{equation}
    where it is understood that the vectors $L_{k}\ket{B}$ and $L_{n}\ket{C}$ are brought into the canonical form \eqref{eq:Virasoro_descendant} using \eqref{eq:Vir_commutator} and the results split into monomial terms of the form 
    $\mathbb{T}\left(\ket{A'},\ket{B'},\ket{C'}\right)$ using the trilinear property of $\mathbb{T}$.
    \item Now the right vector is reduced to a primary one by successive application of 
    \begin{equation}
        \mathbb{T}\left( \ket{h_1},\ket{B},L_{-n}\ket{C}\right)=
        -\sum_{k=-1}^\infty \left({-n+1}\atop{k+1}\right) \mathbb{T}\left( \ket{h_1},L_{k}\ket{B},\ket{C}\right)
    \label{eq:Vir_right}\end{equation}
    (where the formally infinite sum is automatically truncated at the descendant level of $\ket{B}$), and by bringing the vector $L_{k}\ket{B}$ into canonical form using \eqref{eq:Vir_commutator} and using the trilinear property of $\mathbb{T}$ to split the resulting terms into monomial contributions.
    \item Now both the left and right vector are primary, and the middle vector can be simplified using the identity
     \begin{equation}
        \mathbb{T}\left( \ket{h_1},L_{-n}\ket{B},\ket{h_2}\right)=
        -(-1)^n\mathbb{T}\left( \ket{h_1},L_{-1}\ket{B},\ket{h_2}\right)
        +(-1)^n (n-1) h_2 \mathbb{T}\left( \ket{h_1},\ket{B},\ket{h_2}\right)\,,
    \label{eq:Vir_middle}\end{equation}
    for $n\geq 2$ (this condition is necessary to avoid an infinite loop). 
    \item Once the above steps are performed, the matrix element is reduced to the form
    \begin{equation}
        \mathbb{T}\left( \ket{h_1},L_{-1}^m\ket{h_3},\ket{h_2}\right)\,,
    \end{equation}
    which is equal to
    \begin{equation}
        (h_1-h_3-h_2)\dots(h_1-h_3-h_2-m+1)\mathbb{T}\left( \ket{h_1},\ket{h_3},\ket{h_2}\right)\,.
    \end{equation}
    \item
    Finally, the normalisation of the chiral three-point matrices specified in Eq. \eqref{eq:Bnormalisation} means setting
    \begin{equation}
        \mathbb{T}\left( \ket{h_1},\ket{h_3},\ket{h_2}\right)=1\,.
    \end{equation}
\end{enumerate}
Note that each step reduces the matrix element to a combination of ones containing vectors with lower descendant levels. Therefore the computation of the elements of the chiral three-point matrix is best performed by working upwards in the descendant level, and using the already computed matrix elements to terminate the recursion as early as possible.

Before an eventual computation of the chiral three-point matrices using $\mathbb{T}$, however, it is necessary to construct the bases of the chiral spaces $V_h$.

\subsection{Inner products and chiral bases}
While the chiral three-point matrices can be easily computed using basis vectors of the form \eqref{eq:Virasoro_descendant}, these vectors are not orthonormal neither are linearly independent. It is possible to set up an algorithm to compute the scalar product using the conjugation of Virasoro generators $L_n^\dagger=L_{-n}$; however, it turns out to be equally efficient to simply compute scalar products of two vectors $\ket{\Psi}$ and $\ket{\Psi'}$ of the form \eqref{eq:Virasoro_descendant} by using the trilinear function $\mathbb{T}$ as follows:
\begin{equation}
\braket{\Psi}{\Psi'}=\mathbb{T}\left(\ket{\Psi},\ket{0},\ket{\Psi'} \right)\,, 
\end{equation}
with the middle vector chosen as the primary vector with weight $h=0$.

The next problem is to generate a list of basis vectors for the chiral level subspaces $V_h(N)$. This is best done iteratively in the level $N$. Assuming that bases up to $N-1$ have been generated, a list of candidate vectors is created by acting $L_{n-N}$ on $V_h(n)$ with $n=0,\dots,N-1$, keeping only vectors in which the Virasoro operators are ordered as specified in \eqref{eq:Virasoro_descendant}. This ensures that descendants of already eliminated singular vectors are not carried over unnecessarily to the next step, and creates a preliminary list of basis vectors $\{\ket{v_1},\dots,\ket{v_k}\}$. Then the matrix of Virasoro inner product 
\begin{equation}
    g_{ij}=\mathbb{T}\left(\ket{v_i},\ket{0},\ket{v_j}\right)
\end{equation}
of these vectors is computed. The final basis is constructed sequentially from the preliminary list: each vector $\ket{v_i}$ is only added if its inclusion keeps the inner product matrix of the final basis nonsingular, and discarded otherwise. This leads to the construction not only of the basis of $V_h(N)$, but also the matrix of inner products of its elements $g(h,N)$. This becomes an ill-conditioned matrix especially for large $N$, and for computational stability it is advisable to compute its elements, determinant and its inverse (needed later) either exactly using symbolic software such as Mathematica \cite{Mathematica}, or use sufficiently high-precision arithmetic.

\subsection{Gram-Schmidt orthogonalisation and the chiral three-point matrices}
The basis constructed above is not orthonormal, however, this can be solved by Gram-Schmidt orthogonalisation\footnote{In the case of non-unitary CFTs, it is sometimes more convenient to skip this step to avoid the appearance of complex coefficients, which can be compensated by including the inner product in the implementation of the CFTCSA. We refrain from discussing it here to keep the exposition simple. However, it is eventually straightforward to modify the implementation in the main text for this case.}, which (for unitary theories with positive definite inner products) gives a real orthogonal matrix $w(h,N)$ describing the orthonormal basis vectors in terms of the original basis generated above. Again, it is advisable to compute either symbolically or use sufficiently high-precision arithmetic.

The chiral three-point matrices are first computed in the (non-orthonormal) Virasoro basis constructed above:
\begin{equation}
    \tilde{B}^\mathcal{O}\left(h_1,N_1,h_2,N_2\right)_{ij}=g(h_1,N_1)^{-1}_{ik}\mathbb{T}\left(\ket{v_k},\ket{h_\mathcal{O}},\ket{w_j}\right)\,,
\end{equation}
with the vectors $v_k$ and $w_j$ running over the basis of the chiral level spaces $V_{h_1}(N_1)$ and 
$V_{h_2}(N_2)$, respectively. Finally, the chiral three-point matrices needed for the CFTCSA algorithm are obtained by transforming to the orthonormal basis using 
\begin{equation}
    B^\mathcal{O}\left(h_1,N_1,h_2,N_2\right)=
    w\left(h_1,N_1\right)\tilde{B}^\mathcal{O}\left(h_1,N_1,h_2,N_2\right)w\left(h_2,N_2\right)^T\,.
\end{equation}

\section{CFT data for massless boson}\label{sec:boson}

In this appendix we briefly review some basic ingredients of the free bosonic CFT such as its Hilbert space or the computation of vertex operator matrix elements, and also discuss how the matrix elements of creation/annihilation operators can be computed in CFTCSA.

\subsection{Compactified boson field}
Let us begin by the mode expansion of the free bosonic field $\phi$; we impose periodic boundary conditions and compactify the quantum field as $\phi \equiv \phi+2\pi k R$. The normalisation used here for the fundamental bosonic field is slightly different from the one in Subsection~\ref{InhomSGProblem} as 
\begin{equation}
    \phi=\sqrt{4\pi}\varphi\,,\qquad R=\frac{\sqrt{4\pi}}{\beta}\,.
\end{equation}
The mode expansion of $\phi$ can be written as
\begin{equation}
\phi(x,t)=\phi_{0}+\frac{4\pi}{L}\pi_{0}t+\frac{4\pi}{L}\frac{MxR}{2}+i\sum_{k\text{\ensuremath{\neq}0}}\frac{1}{k}\left[a_{k}\exp\left(i\frac{2\pi}{L}k(x-t)\right)+\bar{a}_{k}\exp\left(-i\frac{2\pi}{L}k(x+t)\right)\right]\,.\label{eq:FieldExpansion}
\end{equation}
The winding number operator $M$ has integer eigenvalues and corresponds to the topological charge $Q$. The operators $\phi_{0}$ and $\pi_{0}$ are the zero modes of the field $\phi$ and its conjugate momentum field $\pi=\partial_t \phi$ and  $a_{k}$ and $\bar{a}_{k}$ correspond to left/right oscillator
modes creating/annihilating particles with momenta $p=\pm2\pi|k|/L$. The relevant commutation relations are
\begin{align}
[\phi_{0},\pi_{0}]&=i\,, \nonumber\\
[a_{k},a_{l}]&=[\bar a_k,\bar a_l]=k\delta_{k+l}\,,
\label{eq:FBCCR}\end{align}
that is, $a_{k}$ and $\bar{a}_{k}$ with negative/positive $k$ can be interpreted as creation/annihilation operators. 
Following the usual CFT convention, the free Hamiltonian can be written as 
\begin{equation}
H_{\text{FB}}  =\frac{1}{8\pi}\int_{0}^{L}\text{d}x\left\{:\pi^{2}+\left(\partial_{x}\phi\right)^{2}:\right\}
\label{eq:pcft_FreeBosonHam}
\end{equation}
(where the semicolon denotes normal ordering), which can be recast in the form 
\begin{equation}
H_{\text{FB}}=\frac{2\pi}{L}\left(\pi_{0}^{2}+\frac{R^2M^2}{2}+\sum_{k>0}a_{-k}a_{k}+\sum_{k>0}\bar{a}_{-k}\bar{a}_{k}-\frac{1}{12}\right)\:.\label{HCFT}
\end{equation}
The Virasoro generators of the conformal symmetry algebra are
\begin{equation}
    L_k=\frac{1}{2}\sum_{l=-\infty}^{\infty}:a_{k-l}a_l:\,,\qquad  \bar{L}_k=\frac{1}{2}\sum_{l=-\infty}^{\infty}:\bar{a}_{k-l}\bar{a}_l:\,,
\end{equation}
and we defined the zero modes via
\begin{equation}
    a_0=\pi_0+\frac{RM}{2}\,,\qquad \bar{a}_0=\pi_0-\frac{RM}{2}\,.
\end{equation}
From the above definitions one can easily recover $(L_0+\bar{L}_0)$ in the Hamiltonian \eqref{HCFT}. 

The full chiral algebra of the compactified free boson CFT is the $U(1)\cross U(1)$ current or Kac-Moody algebra, generated by the oscillator modes (including $\pi_0$ and $M$) which determines the structure of the Hilbert space. Continuing to imaginary time $\tau=-it$ and introducing complex coordinates $w=\tau-ix$ and $\bar{w}=\tau+ix$, the $U(1)$ currents are
\begin{equation}
    J(w)=i\partial_w\Phi(w)=\frac{2\pi}{L}\left(\sum_{l=-\infty}^{\infty}a_l e^{-2\pi l w/L}\right)\,,\qquad   \bar{J}(\bar{w})=i\partial_{\bar{w}}\bar{\Phi}(\bar{w})=\frac{2\pi}{L}\left(\sum_{l=-\infty}^{\infty}\bar{a}_l e^{-2\pi l \bar{w}/L}\right)\,,
\end{equation}
and can be easily expressed in terms of mode operators as demonstrated. In the above formulae $\Phi(w)$ and $\bar{\Phi}(\bar{w})$ are the chirally factorised components of the bosonic field.
From $\Phi(w)$ and $\bar{\Phi}(\bar{w})$ the canonical field is simply obtained as  
\begin{equation}
    \phi(w,\bar{w})=\Phi(w)+\bar{\Phi}(\bar{w})\,,
\end{equation}
but one can introduce the dual field $\tilde{\phi}(w,\bar{w})$ as well
\begin{equation}
    \tilde{\phi}(w,\bar{w})=\Phi(w)-\bar{\Phi}(\bar{w})\,.
\end{equation}
The zero mode $\tilde{\phi}_0$ of the dual field is an analogous zero mode that is not present in the mode expansion of the canonical boson field $\phi$; it is the coordinate variable conjugate to the winding number operator $M$ with the commutation relation
\begin{equation}
    \left[\tilde{\phi}_0\,,\frac{MR}{2}\right]=i\,.
\end{equation}
The primary fields of the current algebra are the vertex operators:
\begin{equation}
V_{\nu,\mu}^{\mathrm{\text{cyl}}}(w,\bar{w})=\,:\!e^{i q\Phi(w)+i\bar{q}\bar{\Phi}(\bar{w})}\!:\,,\label{eq:vertexops}
\end{equation}
with the parameters ($\nu,\mu$) ranging over the integers\footnote{In fact, locality of the operator algebra allows another choice as well:  either $\nu \in \mathbb{Z}, \mu \in 2\mathbb{Z}$ or $\nu \in \mathbb{Z}+\frac{1}{2}, \mu \in 2\mathbb{Z}+1$. \cite{1993IJMPA...8.4131K}}, specifying their $U(1)$ charges
\begin{equation}
    q=\frac{\nu}{R}+\frac{R\mu}{2}\,,\qquad \bar{q}=\frac{\nu}{R}-\frac{R\mu}{2}\,,
\end{equation}
whereas their conformal dimensions are given by $h=q^2/2$ and $\bar{h}=\bar{q}^2/2$. 

For simplicity in the following we consider vertex operators only with $q=\bar{q}$, which we relabel as 
\begin{equation}
V_{\nu}^{\mathrm{\text{cyl}}}(w,\bar{w})=\,:\!e^{i \nu\phi(w,\bar{w})/R}\!:\,.
\end{equation}
The normalisation of these vertex operators is specified via the short distance
behaviour of their two-point functions:
\begin{equation}
\langle0|V_{\nu}^{\mathrm{cyl}}(w_{1},\bar{w}_{1})V_{\nu'}^{\mathrm{cyl}}(w_{2},\bar{w}_{2})|0\rangle=\frac{\delta_{\nu,-\nu'}}{|w_{1}-w_{2}|^{4\nu^{2}\Delta}}+\text{subleading terms}\,,\label{eq:vertexop_norm}
\end{equation}
with
\begin{equation}
    \Delta=h=\bar{h}=\frac{1}{2R^2}\,.
\end{equation}
Following Subsection~\ref{subsec:PCFT} and introducing vertex operators $V^{\mathrm{pl}}(z,\bar{z})$ defined on the complex plane, the Hamiltonian of the sine-Gordon model can also be expressed as
\begin{equation}
H_{\text{sG}}=H_{\text{FB}}-\lambda_{\text{sG}}\left(\frac{2\pi}{L}\right)^{2\Delta}\frac{L}{2}\left(V_{+1}^{\mathrm{pl}}(1,1)+V_{-1}^{\mathrm{pl}}(1,1)\right)\delta_{s,s'}\,,\label{HPCFT}
\end{equation}
where $\delta_{s,s'}$ indicates that translation invariance introduces a superselection rule in the conformal spin (c.f. Subsection~\ref{subsec:pert_operator_repr}). Truncating the conformal Hilbert space and using \eqref{HPCFT}, the sine--Gordon TCSA has been implemented numerous times for the investigation of equilibrium \cite{1998PhLB..430..264F,1999NuPhB.540..543F}
and out-of-equilibrium properties of the model \cite{2017PhLB..771..539H,2018PhRvL.121k0402K,2019PhRvA.100a3613H}.

\subsection{Hilbert space of the compact boson CFT}

The Hilbert space of the compact boson CFT is composed of Fock modules built upon the highest weight vectors created from the vacuum state by the vertex operators:
\begin{equation}
    |\nu,\mu\rangle=V^{\text{pl}}_{\nu,\mu}(0,0)|0\rangle=e^{i\nu\phi_0/R+i\mu R \tilde{\phi}_0/2}|0\rangle\,,
\end{equation}
where $|0\rangle$ is the vacuum of the theory. The states $|\nu,\mu\rangle$ have eigenvalues $\nu/R$ under $\pi_0$  and $\mu$ under the winding number operator $M$, that is
\begin{equation}
    \pi_0|\nu,\mu\rangle=\frac{\nu}{R}|\nu,\mu\rangle\,,\qquad M|\nu,\mu\rangle=\mu|\nu,\mu\rangle\,,\qquad \nu,\mu\in\mathbb{Z}\,.
\end{equation}
The vertex operators defined in the previous subsection $V_{\nu}$ correspond to $V_{\nu,0}$ and do not connect modules with different winding numbers. Using the Fock decomposition of the free boson Hilbert space $\mathcal{H}$ we can write 
\begin{equation}
\mathcal{H}=\bigoplus_{_{\nu,\mu}}\mathcal{F}_{\nu,\mu}\,,
\end{equation}
with $\nu,\mu\in\mathbb{Z}$ where each Fock module is spanned by the vectors 
\begin{equation}
a_{-k_{1}}...\,a_{-k_{r}}\bar{a}_{-p_{1}}...\,\bar{a}_{-p_{l}}|\nu,\mu\rangle:\,r,\,l\in\mathbb{N}\:,k_{i}\,,\,p_{j}\in\frac{2\pi}{L}\mathbb{N}^{+}\,,\label{Hilbert}
\end{equation}
which are eigenstates of $H_{\text{FB}}$ with energy 
\begin{equation}
E=\frac{2\pi}{L}\left(\frac{\nu^{2}}{R^2}+\frac{R^2\mu^2}{2}+\sum_{i=1}^{r}k_{i}+\sum_{j=1}^{l}p_{j}-\frac{1}{12}\right)\,.
\label{FockModuleEnergy}
\end{equation}

It is useful to further decompose the Hilbert space into different momentum sectors
as well according to 
\begin{equation}
\mathcal{H}=\bigoplus_{_{\nu,\mu,s}}\mathcal{F}_{\nu,\mu}^{(s)}\,,
\end{equation}
where $\mathcal{F}_{\nu,\mu}^{(s)}$ denotes the $s$ subspace corresponding to conformal spin $s$ (i.e., states with total momentum $2\pi s/L$) of the Fock module spanned by the vectors 
\begin{equation}
a_{-k_{1}}...\,a_{-k_{r}}\bar{a}_{-p_{1}}...\,\bar{a}_{-p_{l}}|\nu,\mu\rangle\,,\quad r,l\in\mathbb{N}\,,\quad k_{i},\,p_{j}\in\frac{2\pi}{L}\mathbb{N}^{+}\,,\quad\sum k_{i}-\sum p_{j}=s\,.
\label{Hilbert2}
\end{equation}
In our numerical studies we used the simplest and most common truncation scheme, when the truncation criterion is the energy\footnote{Other truncation schemes can be applied as well depending on the problem considered \cite{2019PhRvA.100a3613H}.}. In this case we keep states in the truncated conformal Hilbert space whose energy does not exceed $2\pi E_{c}/L$, and so the truncated space is given by 
\begin{equation}
\mathcal{H}_{\mathrm{TCSA}}(E_{c})=\text{span}\left\{ a_{-k_{1}}...\,a_{-k_{r}}\bar{a}_{-p_{1}}...\,\bar{a}_{-p_{l}}|\nu,\mu\rangle:\:\frac{\nu^2}{R^2}+\frac{R^2\mu^2}{2}+\sum_{i=1}^{r}k_{i}+\sum_{j=1}^{l}p_{j}-\frac{1}{12}\leq E_{c}\right\} \,.
\end{equation}
Turning now to the physical operators the matrix element of the operators $a_{k}$ and $\bar{a}_{k}$, can be straightforwardly computed as shortly reviewed. Now we merely mention that these operators act between momentum subspaces of the Fock modules according to
\begin{equation}
\begin{split}a_{k}:\mathcal{F}_{\nu,\mu}^{(s)}\rightarrow\mathcal{F}_{\nu,\mu}^{(s-k)}\,,\\
\bar{a}_{k}:\mathcal{F}_{\nu,\mu}^{(s)}\rightarrow\mathcal{F}_{\nu,\mu}^{(s+k)}\,. & 
\end{split}
\end{equation}
Matrix elements of the vertex operators $V_{\nu}^{\text{pl}}$ can be computed in the conformal basis using the mode expansion of the canonical field $\phi$ (\ref{eq:FieldExpansion}). The $U(1)\times U(1)$ symmetry algebra leads to the superselection rules
\begin{equation}
\begin{split}V_{+\nu'}^{\mathrm{pl}}(0,0):\mathcal{F}_{\nu,\mu}\rightarrow\mathcal{F}_{\nu+\nu',\mu}\,,\\
V_{-\nu'}^{\mathrm{pl}}(0,0):\mathcal{F}_{\nu,\mu}\rightarrow\mathcal{F}_{\nu-\nu',\mu} \,.& 
\end{split}
\end{equation}

\subsection{The computation of vertex operator matrix elements}\label{app:vertexopmatrix}
The matrix elements of vertex operators can be obtained in closed form; here we give explicit formulas developed in the course of the work \cite{1998PhLB..430..264F}. For the sake of brevity, we focus only on $V_{n,0}=V_{n}$ vertex operators. The matrix elements we need are
\begin{equation}
    \langle\Psi'|V_{n}^{\mathrm{pl}}(1,1)|\Psi\rangle\,,
\end{equation}
where $|\Psi'\rangle$ and $|\Psi \rangle$ are two vectors from the. Fock-Hilbert space, that is, they can generally be written as
\begin{equation}
\begin{split}
    |\Psi\rangle =&\frac{1}{N_{\Psi}}\prod_{k=1}^{\infty}a_{-k}^{r_k}\bar{a}_{-k}^{s_k}|\nu,\mu\rangle\\
    |\Psi' \rangle=&\frac{1}{N_{\Psi'}}\prod_{k=1}^{\infty}a_{-k}^{r'_k}\bar{a}_{-k}^{s'_k}|\nu',\mu'\rangle\,,
\end{split}
\end{equation}
where the normalisation
\begin{equation}
N_{\Psi}^2=\prod_{k=1}^{\infty}\langle a_{k}^{r_k}a_{-k}^{r_k}\rangle \langle \bar{a}_{k}^{s_k}\bar{a}_{-k}^{s_k}\rangle=\prod_{k=1}^{\infty}\left(r_k!k^{r_k}\right)\left(s_k!k^{s_k}\right)
\end{equation}
ensures the orthonormality of the conformal basis
\begin{equation}
    \langle\Psi'|\Psi\rangle=\delta_{\nu,\nu'}\delta_{\mu,\mu'}\prod_{k=1}^{\infty}\delta_{r_k,r'_k}\delta_{s_k,s'_k}\,.
\end{equation}
Due to the commutation relations of the oscillator modes and the prescribed normal ordering in the definition of the vertex operators, the vertex fields at position $(1,1)$ can be simply expressed as
\begin{equation}
    V_{n}^{\mathrm{pl}}(1,1)=e^{i\alpha \phi_0}\prod_{k=1}^{\infty} e^{\alpha\frac{a_{-k}}{k}}e^{-\alpha\frac{a_{k}}{k}}e^{\alpha\frac{\bar{a}_{-k}}{k}}e^{-\alpha\frac{\bar{a}_{k}}{k}}\,,
\end{equation}
with
\begin{equation}
    \alpha=\frac{n}{R}\,.
\end{equation}
The matrix element under consideration can now be written as
\begin{equation}
    \begin{split}
      \langle\Psi'|V_{n}^{\mathrm{pl}}(1,1)|\Psi\rangle&=N_{\Psi'}^{-1}N_{\Psi}^{-1}\delta_{\nu',\nu+n}\delta_{\mu',\mu}   \prod_{k=1}^{\infty}\langle a_k^{r'_k} e^{\alpha\frac{a_{-k}}{k}}e^{-\alpha\frac{a_{k}}{k}} a_{-k}^{r_k}\rangle\langle \bar{a}_k^{s'_k} e^{\alpha\frac{\bar{a}_{-k}}{k}}e^{-\alpha\frac{\bar{a}_{k}}{k}} \bar{a}_{-k}^{s_k}\rangle\\
      &=N_{\Psi'}^{-1}N_{\Psi}^{-1}\delta_{\nu',\nu+n}\delta_{\mu',\mu}   \left[\prod_{k=1}^{\infty}\langle a_k^{r'_k} e^{\alpha\frac{a_{-k}}{k}}e^{-\alpha\frac{a_{k}}{k}} a_{-k}^{r_k}\rangle\right]\left[\prod_{k=1}^{\infty}\langle \bar{a}_k^{s'_k} e^{\alpha\frac{\bar{a}_{-k}}{k}}e^{-\alpha\frac{\bar{a}_{k}}{k}} \bar{a}_{-k}^{s_k}\rangle\right]\,.
    \end{split}
    \label{VtxOpFull}
\end{equation}
The chiral factorisation to left and right moving modes emerges in a  manifest way, and so the chiral three-point matrices can be read out simply from this result. The factor $\delta_{\nu',\nu+n}$ is the result of $e^{i\alpha \phi_0}$ acting as a translation operator in the space spanned by the eigenstates of the conjugate variable $\pi_0$ due to the commutation relations \eqref{eq:FBCCR}.

Eq. \eqref{VtxOpFull} shows that the matrix element is factorised into single-mode contributions, which can be evaluated as:
\begin{equation}
   \langle a_k^{r'_k} e^{\alpha\frac{a_{-k}}{k}}e^{-\alpha\frac{a_{k}}{k}} a_{-k}^{r_k}\rangle=\sum_{j'=0}^{\infty} \sum_{j=0}^{\infty} \frac{(-1)^j}{j!j'!}\left(\frac{\alpha}{k}\right)^{j+j'}\langle a_k^{r'_k} a_{-k}^{j'} a_{k}^{j} a_{-k}^{r_k}\rangle\,,
\end{equation}
where
\begin{equation}
    \langle a_k^{r'_k} a_{-k}^{j'} a_{k}^{j} a_{-k}^{r_k}\rangle=k^{j+j'}\binom{r_k}{j}\binom{r'_k}{j'}j!j'! (r_k-j)!k^{r_k-j}\delta_{r_k-j,r'_k-j'}\,.
\end{equation}
Based on the above formulae, the computation of vertex operator matrix elements is straightforward, including their chirally factorised components. Finally, Eq. \eqref{VtxOpFull} shows that the operator $V_n$ only connects Fock modules with $\nu$ indices $\nu'-\nu=n$, and the matrix elements are independent of $\nu$ and $\nu'$. This can be exploited to reduce the memory space required of CFT data.

\subsection{Computing matrix elements of creation/annihilation operators}
\label{AppB4}

In the compactified free boson CFT and its relevant perturbations, the \emph{chiral} creation and annihilation operators are often of particular interest, e.g. for computing  physical observables such as $\partial_x \phi$. The matrices of these operators in the free massless bosons is very sparse, allowing further optimisation well beyond the generic one allowed by chiral factorisation. Here we briefly describe how to implement these optimisations in the framework of CFTCSA, first used in the work \cite{2021arXiv210906869H}.

Using the notations introduced above, we consider the matrix element
\begin{equation}
    \langle \Psi'|a_{-k}|\Psi \rangle\,,\qquad |\Psi \rangle\in  \bigoplus_{_{\nu,\mu}}\mathcal{F}_{\nu,\mu}^{(s)}\,,\qquad |\Psi' \rangle\in  \bigoplus_{_{\nu,\mu}}\mathcal{F}_{\nu,\mu}^{(s+k)}, 
    \label{eq:akmatrixelement}
\end{equation}
where we assumed that $k>0$, in other words, $a_{-k}$ creates a left-moving  chiral boson with momentum $k$. Note that the vectors $|\Psi \rangle$ and $|\Psi' \rangle$ are clearly in different momentum sectors of the Hilbert space. 

We begin by a couple of  general considerations which are not specific to the particular operator:

\begin{enumerate}
    \item The first one regards the use of the Hilbert Space Descriptors in cases when the Hilbert space is composed of more then one fixed momentum sector. This can be achieved in two simple ways:
    \begin{itemize}
        \item It is possible to use a single Hilbert Space Descriptor for the entire Hilbert space, and it may be convenient to organise the various subspaces with consecutive segments in the descriptor corresponding to different momentum sectors. In this case it is necessary to have 
        an additional book-keeping to keep track of the position of different momentum subspaces, which is relatively easy to do.
        \item An alternative solution is to use different Hilbert Space Descriptors for different momentum sectors, arranging them in an array i.e. a Hilbert Space Descriptor List, with each element being the descriptor of the subspace of states with a given fixed momentum.
    \end{itemize}
    Whichever solution is chosen, we assume that we have access to the Hilbert Space Descriptors of the two subspaces $ \bigoplus_{_{\nu,\mu}} \mathcal{F}_{\nu,\mu}^{(s)}$ and $ \bigoplus_{_{\nu,\mu}} \mathcal{F}_{\nu,\mu}^{(s+k)}$ in the standard form described in Subsection \ref{sec:implementation:H-ChDescriptors}, denoting them as $D_{\text{H}}^{(s)}$ and $D_{\text{H}}^{(s+k)}$.
   \item The second general remark concerns the use of the Operator Descriptors and the way the operator action or the matrix elements are computed in the CFTCSA language. We recall that the Operator Descriptor Matrices defined in Subsection \ref{ssec:matrixelem} are matrices with their $(i,j)$th element $k$ indexing the appropriate three-point block of the Operator List. Blocks in the Operator List specify how the operator acts between chiral subspaces. The application of the Matrix Descriptor is very transparent and fast. Nevertheless, there can be other ways to compute matrix elements, which in specific cases can be more convenient and/or better optimised. This is precisely the case for creation and annihilation operators.
\end{enumerate}

For the case of the bosonic mode operators, it fares much better to use a different version of the Operator Descriptors, which we denote by $\hat{D}^{\mathcal{O}}_{\text{Op}}$. This descriptor is not a square matrix, but a $l_{O}\times3$ array where denotes $l_{O}$  the length of the array, and can have a left and right chiral version as well, similarly to its matrix counterpart.  The descriptor $\hat{D}^{\mathcal{O}}_{\text{Op}}$  specifies that the matrix describing the 
mapping from the $j$th chiral subspace to the $i$th chiral subspace is the $k$th element of the Operator List:
\begin{equation}
\hat{D}^{\mathcal{O}}_{\text{Op}}=\begin{pmatrix}1 & 5 & 1\\
2 & 7 & 2\\
3 & 4 & 3\\
\vdots & \vdots & \vdots\\
i & j & k\\
\vdots & \vdots & \vdots
\end{pmatrix}
\end{equation}
Such a descriptor can be imported as a `.dat' file by the function { \tt load( )} in MATLAB as a $l_{O}\times3$ array.

The use of the above Operator Descriptor $\hat{D}^{\mathcal{O}}_{\text{Op}}$ requires an ``inverse Hilbert space descriptor'' abbreviated as merely \emph{Inverse Descriptor} and denoted by $D_{\text{Inv}}$. This descriptor is a list of length $l_\text{Ch}$ with two components and its elements are pairs of integers specifying where the given chiral subspace appears in the Hilbert Space Descriptor. In particular, if the the $k$th element of the inverse descriptor is $(n,m)$, it means, that the $k$th chiral subspace appears at the $n$th position in the Hilbert Space Descriptor as a right subspace and at the $m$th position as a left subspace. If a given chiral subspace is not present in the Hilbert space then $n$ and/or $m=0$. 

As a simple example, assuming that the Hilbert Space Descriptor looks like
\begin{equation}
D_{\text{H}}=\begin{pmatrix}
3 & 1 & 1\\
4 & 2 & 2\\
5 & 3 & 3\\
\vdots & \vdots & \vdots\\
\end{pmatrix}\,,
\end{equation}
the Inverse Descriptor reads as
\begin{equation}
D_{\text{Inv}}=
\begin{pmatrix}0 & 1\\
0 & 2\\
1 & 3\\
\vdots & \vdots
\end{pmatrix}\,.
\end{equation}
For the Inverse Descriptor to be well-defined, it is necessary that each chiral
subspace occurs only once as the left component of a subspace in the Hilbert Space Descriptor. In the case of the free massless boson CFT this is always satisfied for any momentum sector. The Inverse Descriptor can again be imported from a `.dat' file using the function {\tt load( )} in MATLAB as an $l_{\text{Ch}}\times2$
array.

Returning now to our eventual problem of evaluating
\begin{equation}
      \langle \Psi'|a_{-k}|\Psi \rangle\,,
\end{equation}
this task can be easily carried out using the descriptors introduced above. Similarly to the Hilbert Space Descriptors, there are two different Inverse Descriptors for the two subspaces denoted by $D_{\text{Inv}}^{(s+k)}$ and $D_{\text{Inv}}^{(s)}$. The determination of the Operator Descriptor $\hat{D}^{a_{-k}}_{\text{Op}}$ and the Operator List  
$\underrightarrow{\mathcal{B}}^{a_{-k}}$ is straightforward, since the operator $a_{-k}$ cannot change the $U(1)\times U(1)$ charges of the basis vectors, and its action is completely independent of the charges, determined by solely the oscillator mode occupation numbers of the basis vectors. Therefore, the Operator List 
$\underrightarrow{\mathcal{B}}^{a_{-k}}$ consists of matrices solely depending on the descendant levels, and a typical matrix element is easily computed using 
\begin{equation}
    \langle \nu',\mu', \{q,r_q'\}| a_{-k}|\{q,r_q\},\nu, \mu\rangle=\delta_{\nu,\nu'}\delta_{\mu,\mu'}\delta_{r_k'+1,r_k}\sqrt{(r_k+1)k}\,\prod^{\infty}_{\substack{q=1 \\ q \neq k}}\delta_{r_q,r'_q}\,,
\end{equation}
where 
\begin{equation}
    |\{q,r_q\},\nu, \mu\rangle=
    \left(\prod^{\infty}_{q=1}r_q!q^{r_q}\right)^{-1/2}
    \prod^{\infty}_{q=1}a_{-q}^{r_q}|\nu,\mu\rangle\,,
\end{equation}
and the set $\{ q, r_q\}$ incorporates all the occupation numbers $r_q$ for each mode $q$. 
The above formulae encode the fact that the action of the creation operator $a_{-k}$ is only non-zero when the occupation numbers of the modes only differ in the case of the $k$th one, and then only by $1$. As a result, in the conventions introduced above the corresponding Operator List consists of only a few blocks.

With the descriptors introduced above, a generic matrix element \eqref{eq:akmatrixelement} can be evaluated as 
\begin{equation}
\begin{split}
    \langle \Psi'|a_{-k}|\Psi \rangle =\sum_{i=1}^{l_{a_{-k}}}{\vphantom{\sum}}' \mathrm{Tr}
    \Big\{
    &K^{\Psi'}\left(D_{\text{Inv}}^{(s+k)}\left(\hat{D}^{a_{-k}}_{\text{Op}}(i,1),1\right)\right)^{\dagger} 
    \underrightarrow{\mathcal{B}}^{a_{-k}}\left(\hat{D}^{a_{-k}}_{\text{Op}}(i,3)\right)K^{\Psi}\left(D_{\text{Inv}}^{(s)}\left(\hat{D}^{a_{-k}}_{\text{Op}}(i,2),1\right)\right)
    \Big\}\,,
\end{split}
\label{eq:matelem_CreationLeft}
\end{equation}
where we recall that $l_{a_{-k}}$ denotes the length of the Operator Descriptor $\hat{D}^{a_{-k}}$ (which is a $l_{a_{-k}} \times 3$ array) and the primed summation means that we omit the indices $i$ when either $D_{\text{Inv}}^{(s+k)}\left(\hat{D}^{a_{-k}}_{\text{Op}}(i,1),1\right)$ or $D_{\text{Inv}}^{(s)}\left(\hat{D}^{a_{-k}}_{\text{Op}}(i,2),1\right)$ is zero. Note that the main difference from \eqref{eq:matelem_general} is that the summation runs over the new type of Operator Descriptor $\hat{D}^{a_{-k}}$.

Similarly, the action of a mode operator on a vector can be computed as follows  
\begin{equation}
\begin{split}
   K^{a_{-k}\Psi}\left(m\right) =\sum_{i=1}^{l_{a_{-k}}}{\vphantom{\sum}}' 
    &\delta \left[ m,D_{\text{Inv}}^{(s+k)}\left(\hat{D}^{a_{-k}}_{\text{Op}}(i,1),1\right)\right]
    \underrightarrow{\mathcal{B}}^{a_{-k}}\left(\hat{D}^{a_{-k}}_{\text{Op}}(i,3)\right)K^{\Psi}\left(D_{\text{Inv}}^{(s)}\left(\hat{D}^{a_{-k}}_{\text{Op}}(i,2),1\right)\right)
   \,,
\end{split}
\label{eq:Action_CreationLeft}
\end{equation}
where $\delta[a,b]$ denotes the usual Kronecker delta $\delta_{a,b}$. We note that it is necessary to take some precaution when coding so that when $K^{a_{-k}\Psi}\left(m\right)$ zero, it should still be a matrix of appropriate size.

The above formulae are straightforward to implement in higher level programming languages, and are easily extended to annihilation operators or to the mode operators in the other chirality sector.

\bibliographystyle{utphys}
\bibliography{cftcsa}

\end{document}